\documentclass[pra,aps,twocolumn,showpacs,a4paper,10pt,footinbib]{revtex4-1}
\usepackage{amsfonts,amssymb,graphics,graphicx,epsfig,color,times,bbm}
\usepackage[intlimits,centertags,sumlimits]{amsmath}
\usepackage[latin1]{inputenc}
\usepackage{pifont}

\graphicspath{{Bilder/}}

\newcommand{\ket}[1]{\left| #1 \right.\rangle}

\newcommand{\ad}{\hat{a}^\dagger}
\renewcommand{\a}{\hat{a}}
\newcommand{\nb}{\hat{n}}

\newcommand{\nf}{\hat{m}}
\newcommand{\f}{\hat f}
\newcommand{\fd}{\hat{f}^\dagger}

\DeclareMathOperator{\Tr}{Tr}

\newcommand{\Bild}[2][\plotwidth]{\epsfig{file=#2,width=#1}}
\renewcommand{\H}[1]{\hat{H}_{\rm #1}}

\DeclareMathSymbol{\rho}{\mathord}{letters}{"25}
\DeclareMathSymbol{\varrho}{\mathord}{letters}{"1A}

\usepackage[plainpages=false,pdfpagelabels,colorlinks=true,citecolor=blue,pdfdisplaydoctitle=true,pdfpagelayout=TwoPageRight]{hyperref}								
\usepackage{bookmark}	

\usepackage{cancel,ifthen}
\usepackage[normalem]{ulem}

\newcommand{\Komment}[2][NoInPuT]{\ifthenelse{\equal{#1}{NoInPuT}}{}{{\color{blue}\sout{#1}}}{\color{red} #2}}

\begin{document}

\title{The one-dimensional Bose-Fermi-Hubbard model in the ultrafast-fermion limit: Charge density wave phase and MI - CDW phase separation}

\author{Alexander Mering}

\begin{abstract}
In a recent work \cite{Mering2010} we presented results for the Bose-Fermi-Hubbard model (BFHM) in the limit of ultrafast fermions. The present work gives an overview over the used methods and an deeper insight into the implications arising from the treated limit. Starting from the discussion of the phase diagram obtained by numerical means, we develop an analytic theory and derive an effective bosonic Hamiltonian. Arising issues in the Hamiltonian are overcome by inclusion of a back-action, renormalizing the solution of our system. Based on a detailed analysis of the effective Hamiltonian, the phase diagram in the thermodynamic limit is constructed by analytic means and comparison to numerical results obtained by density matrix renormalization group (DMRG) techniques for the full BFHM shows a very reasonable agreement. The most prominent feature of the phase diagram, the existence of a phase separation between Mott insulator (MI) and charge density wave (CDW) is discussed in depth with inclusion of important effects due to the boundary condition. 

\end{abstract}


\date{\today}

\maketitle

\newlength{\plotwidth}
\setlength{\plotwidth}{\columnwidth}
\newcommand{\subplotfactor}{0.6}

\section{Introduction}

Recent experiments on very cold $^4$He \cite{Kim2004} found a supersolid behavior predicted nearly forty years ago \cite{Thouless1969,Andreev1969,Leggett1970}. In a supersolid, superfluidity coexists with a solid structure, where the superfluidity is believed to be related to vacancies in the solid $^4$He with possible issues from remaining $^3$He in the sample as well as disorder (see overview \cite{Balibar2010}). As shown by several authors, supersolids also exist in bosonic systems with non-local interactions \cite{vanOtterlo1995,Batrouni2000,Sengupta2005,Capogrosso-Sansone2010,Mishra2009} or in multi-species systems with a purely local interaction \cite{Buechler2003,Hebert2008,Orth2009,Titvinidze2008,Hubener2009,Keilmann2009,Soeyler2009,Mathey2009,Titvinidze2009}. Beside the prediction of a supersolid phase, a multitude of other phases in mixed systems such as phase separation between the species \cite{Mathey2007,Hebert2007,Batrouni2000,Sengupta2005,Orth2009,Titvinidze2008,Pollet2004}, CDW phases \cite{Mathey2007,Titvinidze2008,Pollet2004,Altman2003,Mering2010} and coexistence regions  of different phases \cite{Hubener2009,Soeyler2009,Mering2010} are reported and are a vital field of research.

As believed in the case of Helium, the supersolid exists because of a particle or hole doping, where the latter one is similar to the mentioned vacancies. For mixtures of bosons and fermions, H\'ebert \emph{et al.} showed by numerical means, that a supersolid of the bosons is only present, if and only if the fermions are at half filling and the bosons are doped away from half filling \cite{Hebert2008}. Special interest gained the situation of double-half filling, where beside the mentioned phases also situations with Luttinger liquid behavior or density wave character \cite{Pollet2006} as well as fermionic CDW in addition to a bosonic CDW of full amplitude \cite{Titvinidze2008} exist.

Here we provide a conclusive analytic theory to understand the physics of the bosonic subsystem in the BFHM for ultrafast fermions. This limit is of natural interest, since in most experimental realizations the fermionic atoms are lighter than the bosonic ones \cite{Guenther2006,Ospelkaus2006}, leading to an increased mobility of the fermions compared to the bosons. Alongside the idea of an effective bosonic theory \cite{Buechler2003} we derive the bosonic Hamiltonian for $J_F\to\infty$, adiabatically eliminating the fermions similar to the approach in \cite{Lutchyn2008}. After explaining the nature of the induced long-range couplings between the bosons, a discussion of the bosonic phase diagram is given together with a study of the influence of boundary effects. All results are accompanied by numerical studies using DMRG for the full BFHM.

So far, no direct connection between the long-range interacting and the mixture case is pointed out to our knowledge, where this similarity can be seen within our effective approach. As shown in \cite{Lutchyn2008,Buechler2003,Buechler2004,Mazzarella2008,Yang2008,Santamore2008,Polak2011}, the inclusion of the second species allows for an effective description of the first species in terms of an effective Hamiltonian. Within linear response theory, the induced interactions for the first species are attractive for mixtures of bosons and fermions \footnote{In \cite{Refael2008} the opposite behavior is reported in contrast to the induced attractive interaction reported in \cite{Lutchyn2008,Buechler2003,Santamore2008,Mering2010}. }, where so far no effects of the long-range density-density interactions are studied in the framework of ultracold atoms, yet. Nevertheless, quantum monte carlo results in two dimensions \cite{Soeyler2009} suggest the appearance of long-range, sign-alternating interactions at least for double half filling similar to our findings.

The framework of our approach is set by the BFHM, describing a mixture of ultracold bosons and fermions in an optical lattice \cite{Albus2003}:
\begin{align}\label{eq:BFHM}
\hat H&=-J_B\sum_j\left(\ad_j\a_{j+1}+\ad_{j+1}\a_{j}\right)+\frac{U}{2}\sum_j\nb_j\left(\nb_j-1\right)\nonumber\\
	&\hspace{0.5cm}-J_F\sum_j\left(\hat c^\dagger_j\hat c_{j+1}+ \hat c^\dagger_{j+1}\hat c_{j}\right)+V\sum_j\nb_j\nf_j,
\end{align}
Here, $\hat a^\dagger, \hat a$ ($\hat c^\dagger,\hat c$) are bosonic (fermionic) creation and annihilation operators and $\nb$ ($\nf$) the corresponding number operators. The bosonic (fermionic) 
hopping amplitude is given by $J_B$ ($J_F$), and $U$ ($V$) accounts for the intra- (inter-) species interaction energy. In the following
we restrict ourselves to the limit of large fermionic hopping, i.e. we assume $J_F\gg U,|V|,J_B$ and the energy scale is set by $U=1$. \\

This work is structured as follows. In section \ref{chap:Friedel}, the dependence of the Mott insulators for vanishing bosonic hopping is studied as a function of the fermionic filling using DMRG. These numerical results lead to the development of an effective bosonic theory described in section \ref{chap:AdiabaticEliminiation} with a detailed discussion of the upcoming coupling constants in section \ref{chap:FreeFermions}. As the couplings display a divergent behavior, an renormalization scheme introducing a free parameter into the system  resolves these unphysical behavior and results in the derivation of a proper effective bosonic Hamiltonian in section \ref{chap:RenormalizationFermions}. Section \ref{chap:SelfconsistentAmplitude} shows possible approaches in the selfconsistent determination of the free parameter from suitable choices of the ground state. Finally, section \ref{chap:ResultsFastFermionsFullModel} discusses the phase diagram in the ultrafast fermion limit with open boundaries as well as in the thermodynamic limit.


\section{Friedel oscillations: fermion induced superpotential}\label{chap:Friedel}

A first, intuitive ansatz to the understanding of the physics in the regime of ultrafast fermions lies in the assumption of a full decoupling of the fermions from the bosons. This uncoupling assumption leads to a homogeneous fermion distribution $\left\langle \nf_j\right\rangle = \rho_F$ and thus to an effective chemical potential for the bosons. The effective potential arising from the the interaction part
\begin{equation}
 V \sum_j \nb_j \nf_j \to V \rho_F\sum_j \nb_j
\end{equation}
simply gives a shift of the bosonic chemical potential as $\mu_B\mapsto \mu_B-V\rho_F$. Nevertheless, this only holds for periodic boundary conditions or in the thermodynamic limit. For open boundaries, substantial in DMRG simulations, the ground state of the fermions is changed in a very important way. Here, the fermionic density displays Friedel oscillations \cite{Friedel1952,Beduerftig1998}, given by 
\begin{equation}
 \left\langle \nf_j\right\rangle =  \frac{N+\frac12}{L+1}-\frac{1}{2(L+1)}\frac{\sin\left(2\pi j\frac{N+\frac12}{L+1}\right)}{\sin\left(\frac{\pi j}{L+1}\right)}.\label{eq:FriedelOscillation}
\end{equation}
Thus, instead of a resulting homogeneous chemical potential $\mu_B$ for the bosons, the system has to be considered as having a site dependent potential $\sum_j \mu_j \nb_j$, where the chemical potential is given by $\mu_j =\mu_B- V   \left\langle \nf_j \right\rangle$. This site dependent chemical potential introduces a qualitatively new feature to the system which is equivalent to the disordered Bose-Hubbard model (dBHM), respectively the BHM with a superpotential. For the MI phases, it is well justified to neglect any possible influence of the bosons onto the fermions even for finite but large values of $J_F$. At other densities it will turn out that this does not hold. For this reason, we only discuss the Mott lobes at this point and map out the dependence of the MI boundaries as function of the fermionic density $\rho_F$ for $J_B=0$.

On ground of this superpotenial BHM we are now able to discuss the phase diagram for $J_B=0$ in a straightforward way. Considering particle-hole excitations \cite{Freericks1996}, we find the chemical potentials for the upper and lower lobe of the $n-$th Mott insulator to satisfy
  \begin{align}
   \mu_n^+ &= V n \rho_F + V \min_j \left\langle \nf_j \right\rangle,\\
    \mu_n^- &= V n \rho_F + V \max_j \left\langle \nf_j \right\rangle.
  \end{align}
  
  \begin{figure}[t]
  \centering
   \Bild{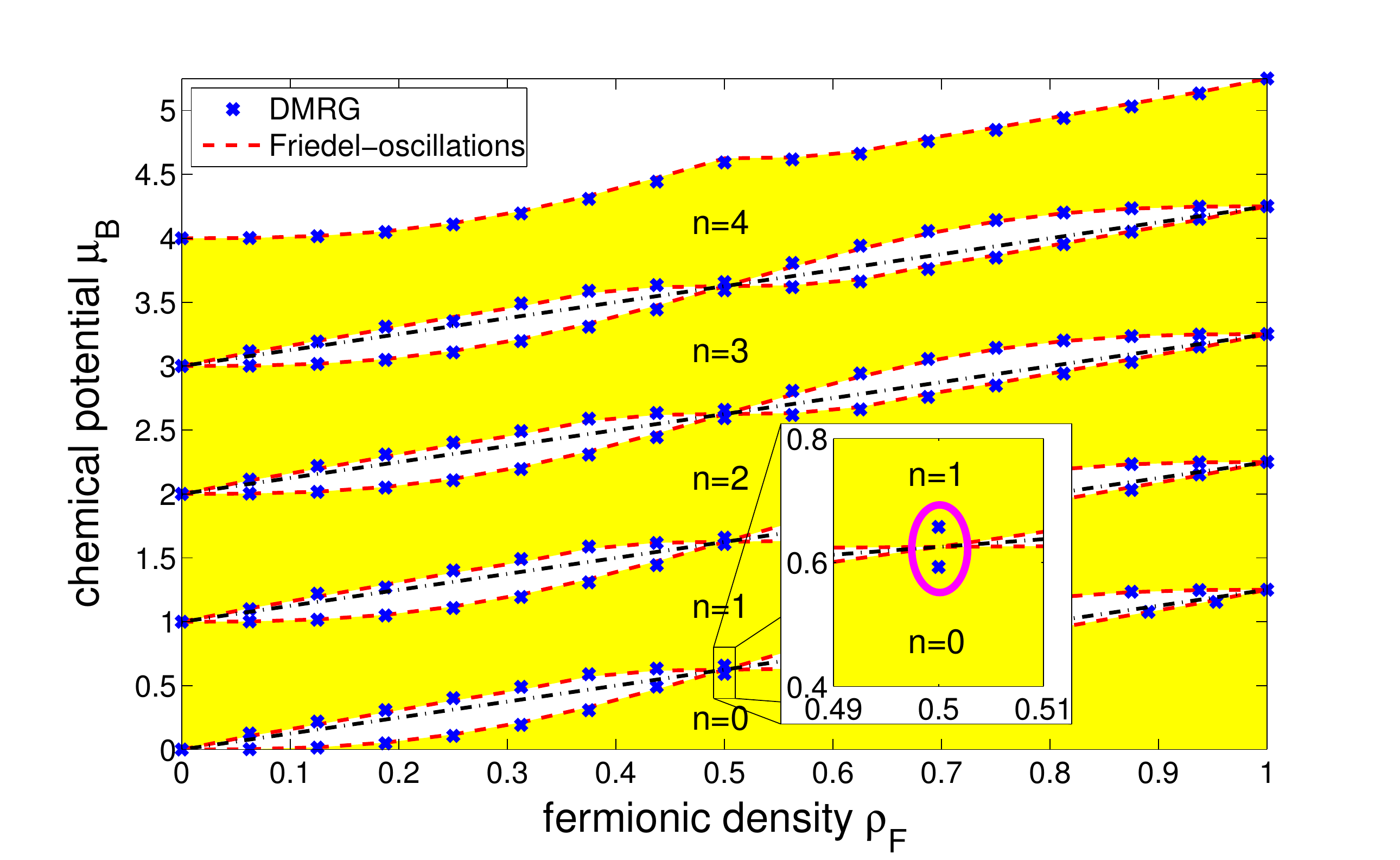}
   \caption{(Color online) Phase diagram of the BFHM for zero bosonic hopping $J_B=0$ as a function of the fermionic filling $\rho_F$ at large hopping $J_F=10$. The yellow shaded regions each represent the different Mott lobes, where the shrinking of the Mott lobes is a typical feature of the underlying effective potential as described in the main text. The numerical data are obtained for $L=64$ with $V=1.25$ and the agreement with the analytic prediction from the Friedel oscillations (dashed line along the data) is very good. The mean-field shift $V\rho_F$ is indicated by the straight dash-dotted lines.  The zoom indicates the non-closing of the Mott lobes for half fermionic filling as discussed in the main text.}
   \label{fig:ZeroHoppingFriedelOscillation}
    \end{figure}
  
  This result is shown in figure \ref{fig:ZeroHoppingFriedelOscillation}, where numerical results as well as the corresponding analytic curves are presented. In the figure it may be recognized, that with increasing fermion density, the Mott insulators are first shrinking, opening a gap between two adjacent Mott insulators. This gap is maximal around quarter filling with a reclosing for half filling. Beyond half filling, the same structure arises due to particle hole symmetry of the fermions. This property is superimposed onto a mean-field shift $V\rho_F$ which comes from the first term in equation \eqref{eq:FriedelOscillation}.  Although the explanation for the phase diagram is quite intuitive, it lacks an important feature. As can be seen from the zoom in figure \ref{fig:ZeroHoppingFriedelOscillation}, the Mott insulators do not exactly close the gap for half fermionic filling $\rho_F=\frac12$. This behavior cannot be understood from the Friedel oscillations, since for half filling these are in phase with the lattice spacing, i.e., the fermionic density is constant in this case. This observation is the starting point of our study of the phase diagram with special focus on half fermionic filling, i.e., $\rho_F=1/2$.
  
\begin{figure}[t]
  \centering
   \Bild[\plotwidth]{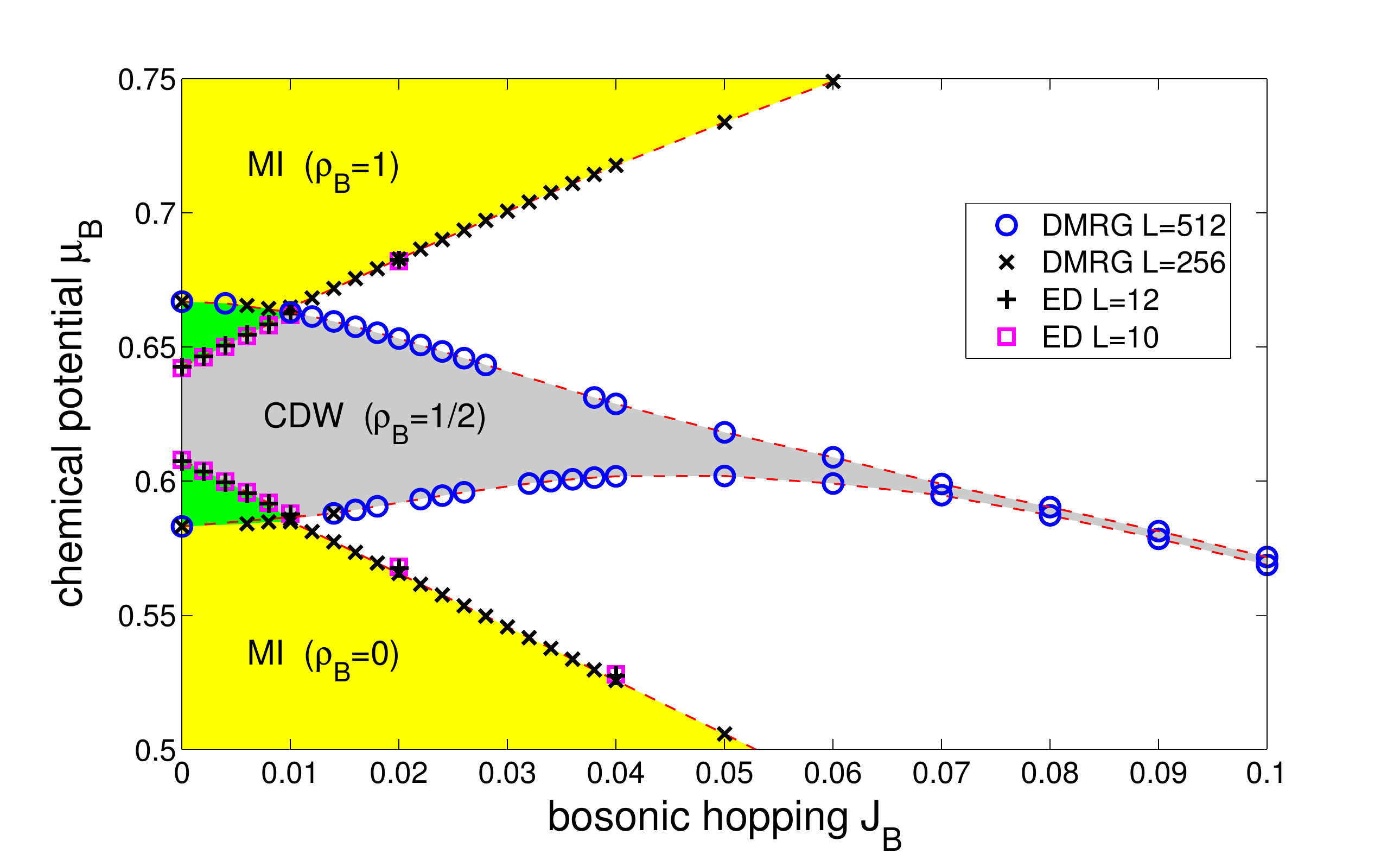}
  \caption{(Color online) Phase diagram of the BFHM for $\rho_F=\frac12$. Beside the expected Mott insulating lobes (yellow), an incompressible CDW for half filling is found (gray). Most prominent feature is the overlap between the CDW phase and each of the MI (green), indicating a thermodynamic instable region as discussed in the main text.  The numerical data (points) were obtained for $V=1.25$ and $J_F=10$, using DMRG and ED for small lattices with system sizes as indicated in the legend. The lines are to guide the eye.}
   \label{fig:PhaseDiagramFastFermions}
    \end{figure}

  As a first step we construct the phase diagram for the lowest two lobes by numerical means using DMRG and exact diagonalization (ED). As shown in figure \ref{fig:PhaseDiagramFastFermions}, beside the usual Mott lobes, a third incompressible phase can be observed. Whilst the Mott lobes do not touch each other, opening a gap between them, a CDW phase extends even beyond this gap, partially overlapping with the MI.
  This region of metastability  indicates the existence of a thermodynamic instable phase with coexistence of Mott insulator and CDW and both, the existence and the extent can be fully understood by an effective theory as pointed out in the upcoming sections.
  
\section{Adiabatic elimination of the fermions}\label{chap:AdiabaticEliminiation}

\begin{figure}
  \centering
\Bild{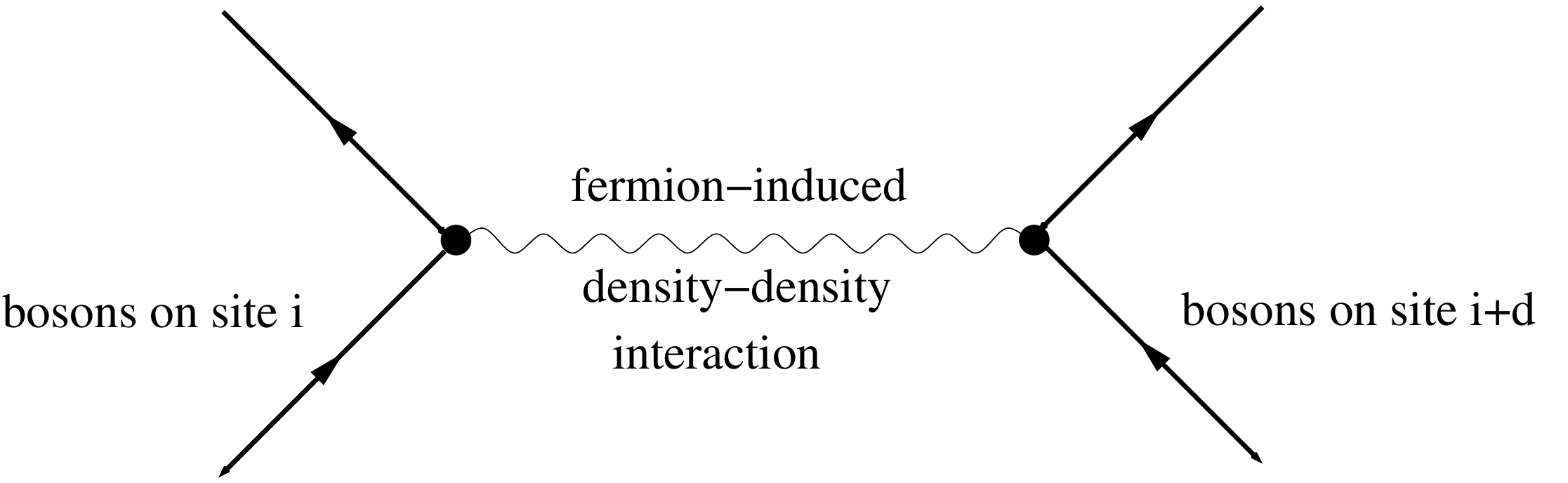}
 \caption{Feynman graph representing the fermion induced density-density interaction.}\label{fig:FeynmanDensDens}
\end{figure}

In order to understand the presented phase diagram we derive an effective bosonic capturing all physical effects. From a simple rewriting, we split the full Hamiltonian \eqref{eq:BFHM} into a bosonic part $\H B$, a fermionic part $\H F$ and an interaction part $\H I$, i.e., $\hat H = \H B + \H F + \H I$, with
\begin{align}
 \H B &= -J_B\sum_j\left(\ad_j\a_{j+1}+\ad_{j+1}\a_{j}\right)+\frac{U}{2}\sum_j\nb_j\left(\nb_j-1\right)\label{eq:BosonicHamiltonian}\\
   \H F &= -J_F\sum_j\left(\hat c^\dagger_j \hat c_{j+1}+ \hat c^\dagger_{j+1} \hat c_{j}\right)+ V \sum_j \widetilde n_j \nf_j\label{eq:FermionicHamiltonian}\\
 \H I &= V\sum_j(\nb_j-\widetilde n_j)\nf_j.
\end{align}
At this step, we already introduced a bosonic mean-field potential $\widetilde n_j$ in $\H I$. This term serves later in the renormalization as discussed in section \ref{chap:RenormalizationFermions}. For the moment, this term is kept for simplicity, without a deeper meaning. The effective bosonic Hamiltonian is found from an adiabatic elimination, which is performed in the framework of the scattering matrix 
\begin{equation}
 \hat{\mathcal S} ={ \mathcal T}\exp\left\{-\frac i\hbar \int_{-\infty}^\infty {\rm d}\tau  \H I(\tau)\right\} \label{eq:SMatrix}
\end{equation}
of the full system in the interaction picture, i.e., $\H {I}(\tau) = e^{-\frac{i}{\hbar}(\H B +\H F)\tau}\  \H I\  e^{\frac{i}{\hbar} (\H B + \H F)\tau}$ and $\mathcal T$ being the time ordering operator. Tracing over the fermionic degrees of freedom yields the bosonic scattering matrix via $ \hat{\mathcal S} ^{\rm B}_{\rm eff} = \Tr_{\rm F} \hat{\mathcal S}$. At this point, we make use of the so-called \emph{cumulant expansion} \cite{Gardiner1985,Kubo1962}, which relates the average of an exponential $\left \langle \exp \{ s X \} \right\rangle_{\rm X}$ (with respect to a stochastic variable $X$) to the exponential of the averages, i.e., the higher order cumulants of the stochastic variable
\begin{equation}
 \left \langle \exp \{ s X \} \right\rangle_{\rm X} = \exp \left\{\sum_{m=1}^\infty \frac{s^m}{m!} \langle\langle X^m\rangle\rangle \right\}.\label{eq:CumulantExpansion}
\end{equation}
Since the cumulants for the fermionic system vanish for orders higher than two due to the nature of the fermionic state, the final expression for the bosonic S-matrix is given by
\begin{align}
   \hat{\mathcal S} ^{\rm B}_{\rm eff} &= \mathcal T \exp \Biggl\{  -\frac i\hbar\   V \sum_j \int_{-\infty}^\infty {\rm d}\tau\   \Bigl(\nb_j(\tau)-\widetilde n_j\Bigr) \langle\langle\nf_j(\tau)\rangle\rangle_{\rm F}\notag\\
&\hspace{-0.2cm} +\frac12 \left(-\frac{i}{\hbar}\right)^2V^2 \sum_{j,l} \int_{-\infty}^\infty {\rm d}\tau\    \int_{-\infty}^\infty {\rm d}\sigma\times\\
&\hspace{0.2cm}\times\Bigl(\nb_j(\tau)-\widetilde n_j\Bigr)  \Bigl(\nb_l(\sigma)-\widetilde n_l\Bigr)  \langle\langle\mathcal T\ \nf_j(\tau)\nf_l(\sigma)\rangle\rangle_{\rm F} \Biggr\}\notag.
 \end{align}
At this point, the time ordering in the fermionic cumulants is important as discussed in \cite{Kubo1962,Kubo1963}. So far, no approximations are used, thus the effective bosonic S-matrix is exact. To derive an effective Hamiltonian for the bosonic system we apply a Markov approximation \cite{Louisell1973,Carmichael1993}, replacing  the two-time bosonic density-density operators by equal time operators, i.e.,
 \begin{align}
  &\iint {\rm d}\tau {\rm d}\sigma\       \Bigl(\nb_j(\tau)-\widetilde n_j\Bigr)  \Bigl( \nb_l(\sigma)  -\widetilde n_l\Bigr) \langle\langle\mathcal T\ \nf_j(\tau)\nf_l(\sigma)\rangle\rangle_{\rm F}\notag \\
 &\hspace{1cm}\mapsto   \int {\rm d}\tau\     \Bigl(\nb_j(\tau)-\widetilde n_j\Bigr)  \Bigl( \nb_l(\tau)  -\widetilde n_l\Bigr)\times\label{eq:MarkovApproximation}\\
&\hspace{2cm}\times\int {\rm d}\sigma\     \langle\langle\mathcal T\ \nf_j(\tau)\nf_l(\sigma)\rangle\rangle_{\rm F} . \notag
\end{align}
This approximation is valid since the timescale of the fermionic system is $1/J_F$ and therefore much shorter than any other timescale in the system. Using the Markov approximation, we rewrite the bosonic S-matrix as
\begin{equation}
 \hat{\mathcal S}_{\rm eff}^B ={ \mathcal T}\exp\left\{-\frac i\hbar \int_{-\infty}^\infty {\rm d}\tau  H_I^{\rm eff}(\tau)\right\},
\end{equation}
which defines the effective bosonic interaction Hamiltonian in the interaction picture
\begin{equation}
 \begin{split}
   \H{I}^{\rm eff}(\tau) &=  V \sum_j \Bigl(\nb_j(\tau)-\widetilde n_j\Bigr) \langle\langle\nf_j(\tau)\rangle\rangle_{\rm F}\\
&\hspace{-0.5cm}- i \frac{V^2}{2 \hbar} \sum_{jl}    \Bigl(\nb_j(\tau)-\widetilde n_j\Bigr)  \Bigl(\nb_l(\tau)-\widetilde n_l\Bigr)\times\\
&\hspace{1cm}\times\int_{-\infty}^\infty {\rm d}\sigma\   \langle\langle\mathcal T\ \nf_j(\tau)\nf_l(\sigma)\rangle\rangle_{\rm F}.
 \end{split}
\end{equation}
The first order cumulant $\langle\langle\nf_j(\tau)\rangle\rangle_{\rm F}$ is equal to the expectation value and the second order cumulant, given by $\langle\langle \nf_j(\tau)\nf_l(\sigma)\rangle\rangle=\langle \nf_j(\tau)\nf_l(\sigma)\rangle-\langle\nf_j(\tau)\rangle\langle\nf_l(\sigma)\rangle$ only depends on the difference $T$ of the times $\tau$ and $\sigma$ and the distance $d$ between the  sites $j$ and $l$ as will turn out in the next section. The final form of the effective bosonic Hamiltonian in the Schr\"odinger picture can thus be written as
\begin{align}
  \H{I}^{\rm eff} &=   -J_B\sum_j\left(\ad_j\a_{j+1}+\ad_{j+1}\a_{j}\right)+\frac{U}{2}\sum_j\nb_j\left(\nb_j-1\right)\notag\\
 &\hspace{1cm}+ V \sum_j \Bigl(\nb_j-\widetilde n_j\Bigr) \langle\nf_j\rangle_{\rm F}\label{eq:effectiveBHMFull}\\
  &\hspace{1cm}+\sum_{j}\sum_{d=-\infty}^\infty   g_d(\rho_F) \Bigl(\nb_j-\widetilde n_j\Bigr)  \Bigl(\nb_{j+d}-\widetilde n_{j+d}\Bigr).\notag
\end{align}
This result allows to distinguish the effect of the fermions on the bosonic subsystem into two cases: (i) the mean-field interaction (1st order) and (ii) the induced density-density interactions (2nd order). Physically, in second order, the fermions act as virtual photons (see figure \ref{fig:FeynmanDensDens}), inducing long range interaction between the bosons given by the coupling constants 
\begin{equation}
 g_d(\rho_F) = - i \frac{V^2}{2 \hbar} \int_{-\infty}^\infty {\rm d}T\   \langle\langle\mathcal T\ \nf_j(T)\nf_{j+d}(0)\rangle\rangle_{\rm F}. \label{eq:CouplingsGeneral}
\end{equation}

As can be seen from the couplings, the remaining task in the calculation of the effective Hamiltonian is to calculate the fermionic density-density correlator  $\langle\langle\mathcal T\ \nf_j(T)\nf_{j+d}(0)\rangle\rangle_{\rm F}$ using the fermionic Hamiltonian in (\ref{eq:FermionicHamiltonian}). Subsequently, the phase diagram can be constructed.

\section{Couplings \texorpdfstring{$g_d(\rho_F)$}{gd(rhoF)} for free fermions}\label{chap:FreeFermions}

The calculation of the couplings relies on the straightforward solution of the fermionic problem given by \eqref{eq:FermionicHamiltonian}. In the present section we restrict ourselves on the case of free fermions, i.e., $\widetilde n_j\equiv0$, assuming a full decoupling of bosonic and fermionic quantities.

Assuming the ground state of the fermionic subsystem to be the ground state of the free fermionic Hamiltonian
\begin{equation}
 \H F = -J_F\sum_j\left(\hat c^\dagger_j\hat c_{j+1}+ \hat c^\dagger_{j+1}\hat c_{j}\right),
\end{equation}
the local density $\langle\nf_j\rangle_{\rm F}$ and the couplings $ g_d(\rho_F) $ are easily calculated in momentum space. Applying a Fourier transform
\begin{equation}
  \hat c_j = \frac{1}{\sqrt{L}}\sum_{k=-\frac L2}^{\frac L2-1} e^{-2\pi i\frac{kj}{L}} \f_k,\label{eq:FourierTrafo}
\end{equation}
the fermionic Hamiltonian transforms into 
\begin{equation}
 \H F = -2 J_F\sum_k  \cos ( 2\pi \frac kL)\ \fd_k\f_k,
\end{equation}
and the ground state is given by the Fermi sphere $\mathcal K_F= \{ k |\  |k| \le k_F\}$ with Fermi momentum $k_F=N_F/2$. Here $N_F$ is the number of fermions in the system and $L$ is the number of sites. 
The real-space density operator $\nf_j(\tau)$ in the interaction picture and in Fourier space is given by
\begin{align}
 \nf_j(\tau)&=\frac{1}{L}\sum_{k_1,k_2} e^{-\frac i\hbar \tau 2J_F\left[ \cos(2\pi\frac {k_1}L)-\cos(2\pi\frac {k_2}L)\right]} \times\\
 &\hspace{1cm}\times e^{-2\pi i\frac{(k_1-k_2)j}{L}} \fd_{k_1}\f_{k_2},\notag
\end{align}
which together with the ground state $ \left|\Psi_F\right\rangle = \prod_{k\in \mathcal K_F} \fd_k \left|0\right\rangle$ and the four-point function
\begin{equation}
 \begin{split}
\langle \fd_{k_1}\f_{k_2} \fd_{k_1^\prime}\f_{k_2^\prime}\rangle &= \delta_{k_1,k_2}\delta_{k_1^\prime ,k_2^\prime}\Theta(k_F-|k_1|) \Theta(k_F-|k_1^\prime|) \\
&+\delta_{k_1,k_2^\prime}\delta_{k_1^\prime, k_2}\Theta(k_F-|k_1|) \Theta(|k_1^\prime|-k_F),  
 \end{split}
\end{equation}
allows to directly calculate the density-density cumulant:
\begin{align}
 \langle\langle\nf_j(T)\nf_{j+d}(0)\rangle\rangle_{\rm F} &= \frac{1}{L^2}\sum_{k_1\in \mathcal K_F}\sum_{k_1^\prime\not\in \mathcal K_F}\\
 &\hspace{-2.5cm} e^{-\frac i\hbar 2J_F T \cos(2\pi\frac{k_1}L)}e^{\frac i\hbar 2J_FT\cos(2\pi\frac{k_2}L)} e^{-{2\pi i} \frac{d k_1}L}e^{{2\pi i}\frac {dk_1^\prime}L}\notag.
\end{align}

\begin{figure}
 \centering
 \Bild{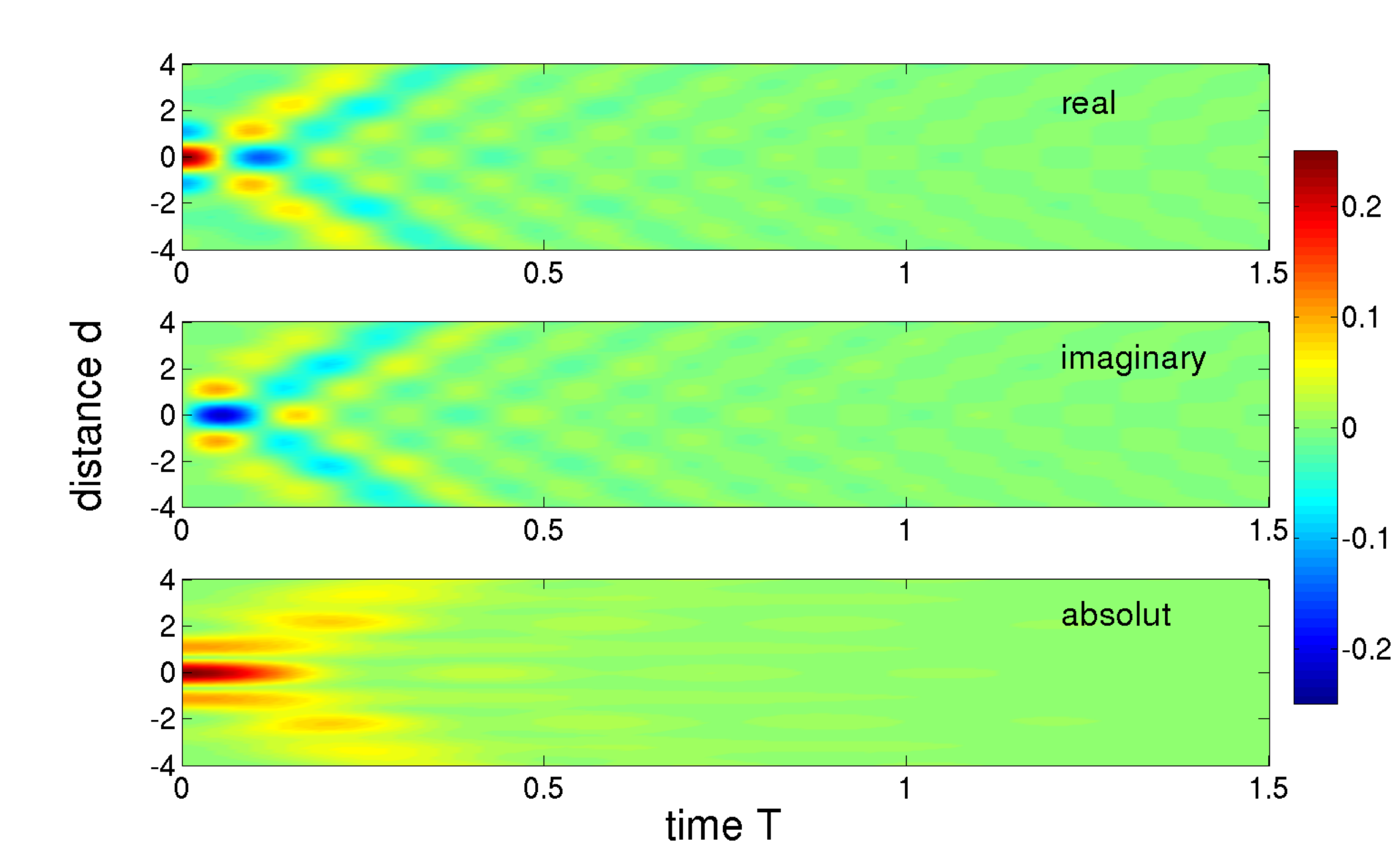}
 \caption{(Color online) Density-density cumulant \eqref{eq:CumulantFreeFermions} for free fermions split in real and imaginary parts as well as the absolute value. $\rho_F=1/2$ and $J_F=10$. The sharp peak around $T=0$ shows the validity of the used Markov approximation.}
 \label{fig:DensityDensityCumulant}
\end{figure}

This is done for $T>0$, where time ordering is irrelevant. To simplify the calculation of the momentum sums it is more convenient to switch to the thermodynamic limit $L\to\infty$ which is reached by defining  $\xi=\frac kL$ and changing $\frac 1L\sum_k$ to $\int {\rm d}\xi$. Together with a further substitution $2\pi \xi \mapsto \xi$, the cumulant simplifies to
 \begin{equation}
  \begin{split}
     \langle\langle\nf_j(T)\nf_{j+d}(0)\rangle\rangle_{\rm F} &=  \frac{1}{\pi^2} \int_ {0}^{\rho_F\pi} {\rm d} \xi \int_ {\rho_F \pi}^{\pi} {\rm d} \xi^\prime\ \\
  &\hspace{-2cm}\cos (d\xi)\cos(d\xi^\prime )\ e^{-\frac i\hbar 2 J_FT   \left[ \cos(\xi)-\cos(\xi^\prime)\right]},
  \end{split}\label{eq:CumulantFreeFermions}
 \end{equation}
which is it's final form. The applicability of the made Markov approximation \eqref{eq:MarkovApproximation} can be seen in figure \ref{fig:DensityDensityCumulant}, where the real, imaginary and absolute values of the cumulant show a sharp localization around $T=0$. 
 
 The knowledge of the cumulant allows to perform the time-integration in \eqref{eq:CouplingsGeneral}. Using time-symmetry and the Riemann-Lebesgue lemma \cite{Bochner1949}, the final result for the coupling constants for the case of free fermions is given by
\begin{equation}
 g_d(\rho_F) = -\frac{V^2}{2\pi^2J_F}  \int_ {0}^{\rho_F\pi} {\rm d} \xi \int_ {\rho_F \pi}^{\pi} {\rm d} \xi^\prime\   \frac {\cos (d\xi)\cos(d\xi^\prime )}{ \cos(\xi)-\cos(\xi^\prime)}. \label{eq:CouplingsFreeFermions}
\end{equation}

Prior to our discussion of the phase diagram, several important properties of the arising coupling constants have to be discussed. The first thing to observe is the existence of a particle-hole symmetry $g_d(\rho_F)=g_d(1-\rho_F)$ which can be shown by substituting $\xi \to \pi-\xi$ and $\xi^\prime \to \pi-\xi^\prime$ and interchanging $\xi \leftrightarrow \xi^\prime$ afterwards. This is a natural consequence of the underlying fermionic system. 

\begin{figure}[t] 
  \centering
   \Bild{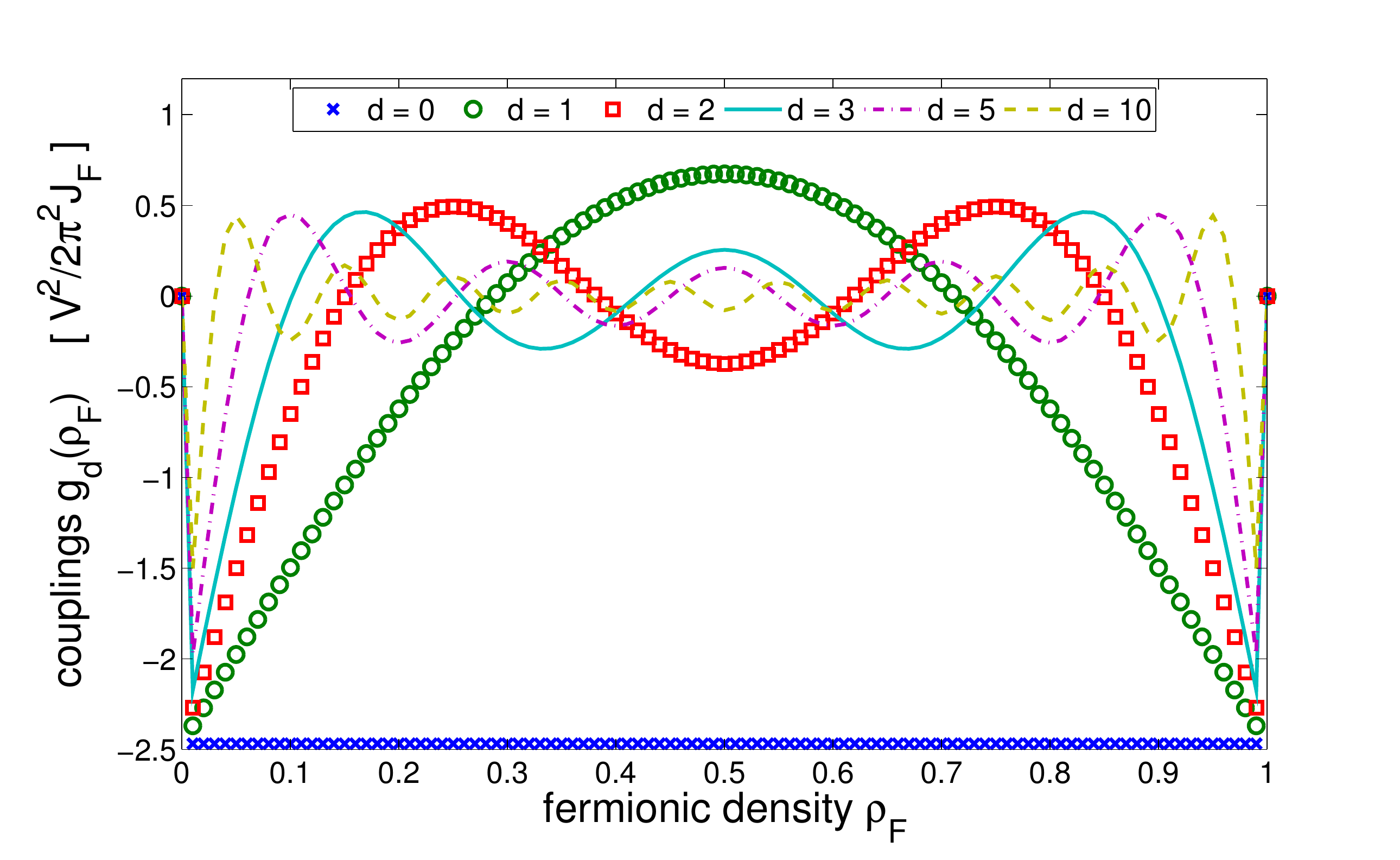}
   \caption{(Color online) Dependence of the coupling strength $ g_d(\rho_F)$ for various distances $d$ on the fermionic filling $\rho_F$. One can obviously see the particle-hole symmetry, reflecting $g_d(\rho_F)=g_d(1-\rho_F)$ as well as the singular behavior for integer filling.}
   \label{fig:CouplingsRealSpaceRhoFree}
    \end{figure}

\begin{figure}[t]  
  \centering
   \Bild[1.15\plotwidth]{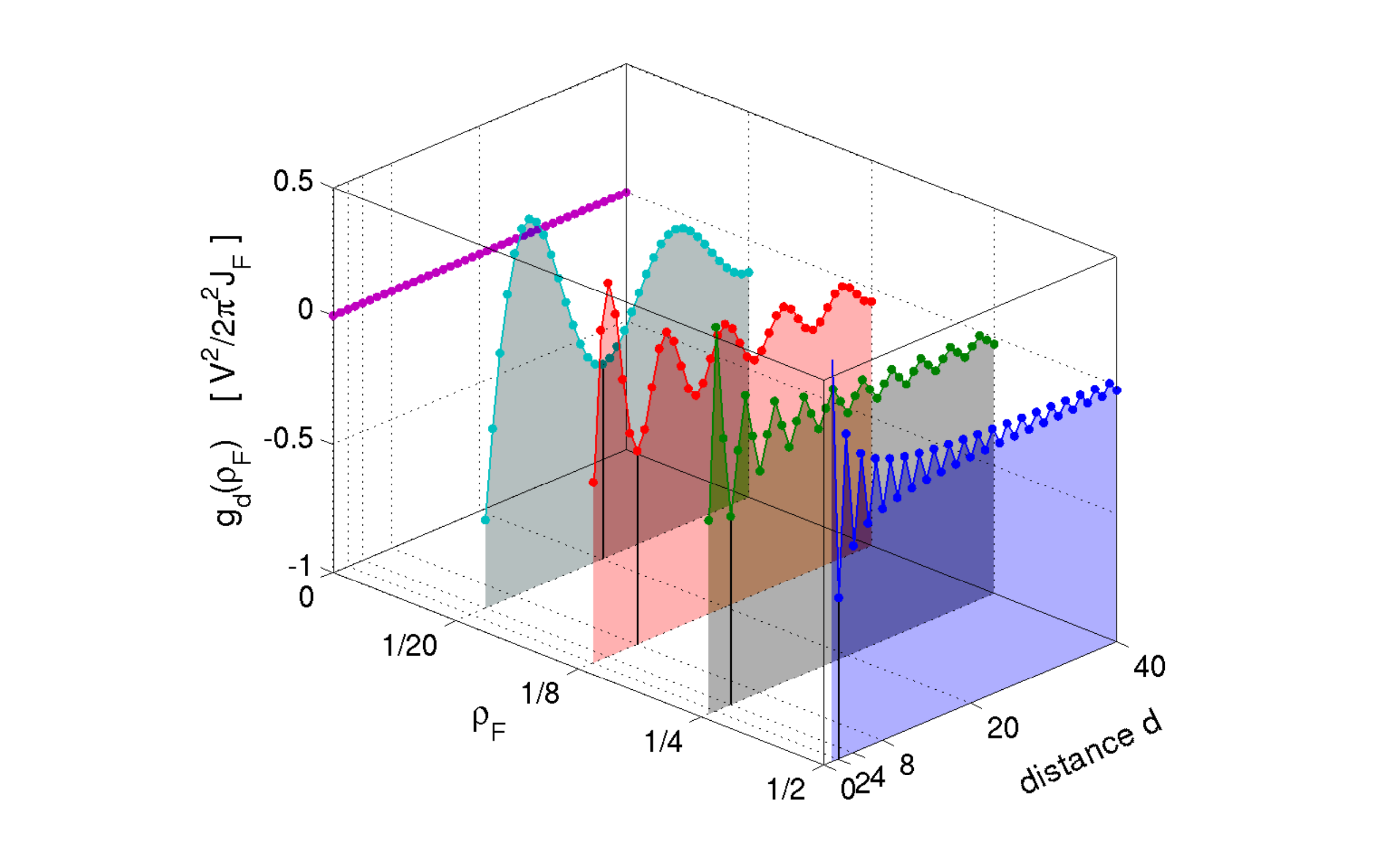}
   \caption{(Color online) Dependence of the coupling strength $ g_d(\rho_F)$ for selected densities $\rho_F = 0, 1/20, 1/8, 1/4, 1/2$ on the distance $d$. The periods of the oscillations are $1/\rho_F = \infty, 20, 8, 4, 2$. For all cases, the signs in the minima are negative and the maxima positive with a strict alternation from site to site for the case of half filling.}
   \label{fig:CouplingsRealSpaceDistanceFree}
    \end{figure}

Secondly, for any density $\rho_F\not=0,1$, the local interaction is reduced, i.e.,
\begin{equation}
g_0(\rho_F)\Bigr|_{\rho_F\not=0,1} =  -\frac{V^2}{8J_F}<0.
  \end{equation}
 This negative shift is in full agreement with the results from \cite{Lutchyn2008,Tewari2009,Buechler2003}, predicting the enhancement of the superfluid phase because of a reduction of the on-site interaction $U$ of the bosons. Beyond this simple local renormalization, we are able to incorporate further interaction effects affecting the phase diagram in this regime. A detailed discussion of the coupling constants nevertheless reveals some important issues to be overcome.  Figure \ref{fig:CouplingsRealSpaceRhoFree} shows the numerical results for the coupling constants as a function of the fermionic filling $\rho_F$. For the case of zero or unity fermionic filling it should be mentioned, that the coupling constants are zero in these two cases, whereas the limit $\lim_{\rho_F \to 0,1} g_d(\rho_F)=g_0(\rho_F)$ is unequal to zero. Figure \ref{fig:CouplingsRealSpaceDistanceFree} shows the dependence of the couplings on the distance $d$ for selected densities $\rho_F$.  One can see a periodic modulation of the couplings,  with the wavelength of the modulation given by $1/\rho_F$ (for $\rho_F<\frac12$, otherwise the wavelength is given by $1/(1-\rho_F)$). This behavior of the couplings is typical for induced couplings of the RKKY-type (Rudermann-Kittel-Kasuya-Yosida) \cite{Rudermann1954,Kasuya1956,Yosida1957}. The most interesting case can be found for $\rho_F=1/2$. In this case, the wavelength of $2$ leads to a strict alternation in the sign of the couplings from site to site. As a result, the effective Hamiltonian (\ref{eq:effectiveBHMFull}) displays repulsive nearest-neighbor, attractive next-nearest-neighbor, repulsive next-next-nearest-neighbor interaction and so on. See \cite{Soeyler2009} for a similar, numerical study in this case for two dimensions.\\
  
\begin{figure}[t]
  \centering
   \Bild[0.95\plotwidth]{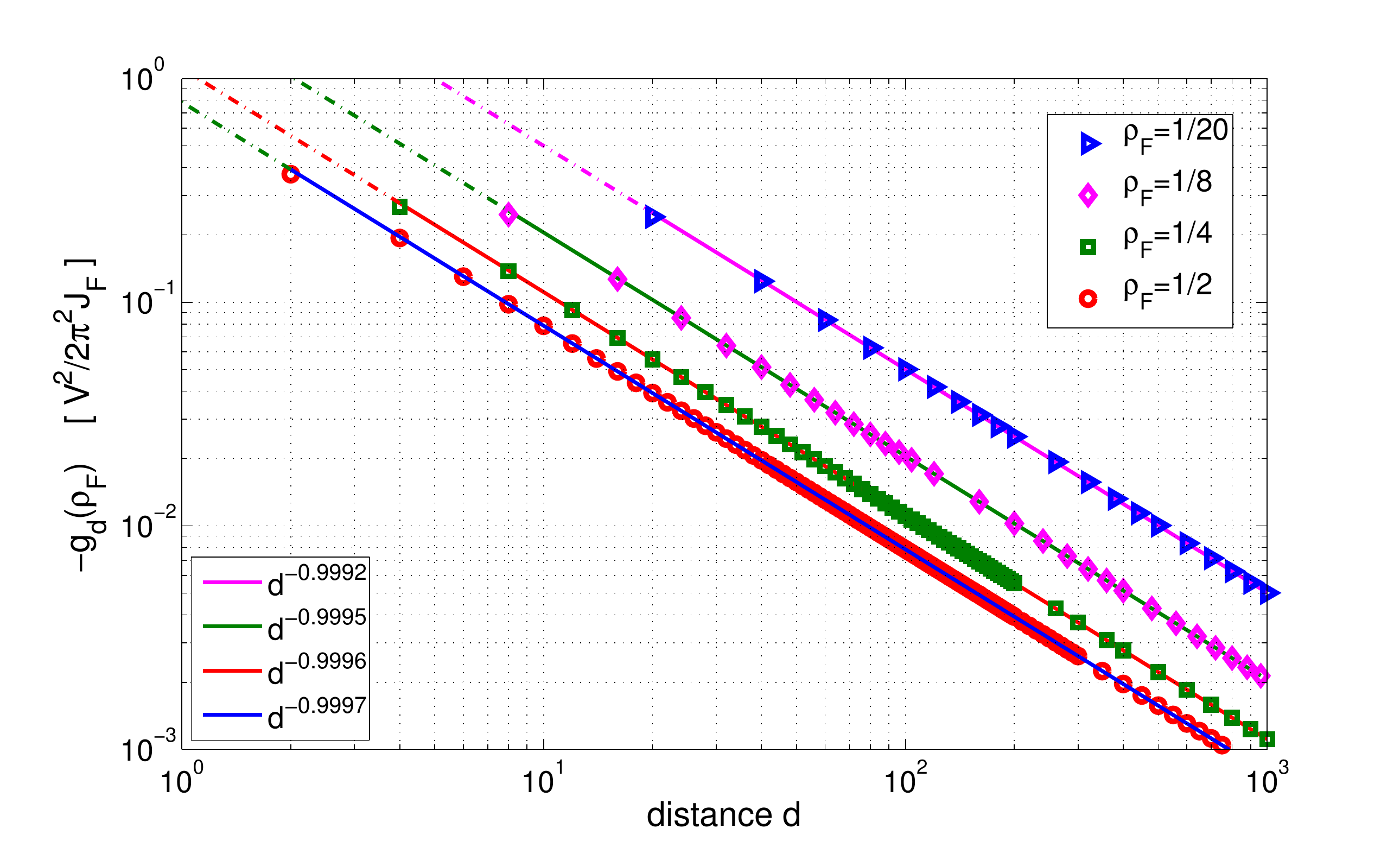}
   \caption{(Color online) Absolute value of the couplings $g_{\frac{m}{\rho_F}}(\rho_F)$ as a function of distance $d$ for selected densities $\rho_F$. Points are the numerical integration of the double integral and the solid lines are a linear fit in the double logarithmic plot. As indicated, the fitting yields a decay of the couplings inverse to the distance for all densities $\rho_F$. The slight deviation of the exponent from one can be attributed to the limited set of fitting points.}
   \label{fig:CouplingsRealSpaceHalfFree}
    \end{figure}

Figure \ref{fig:CouplingsRealSpaceHalfFree} directly reveals the mentioned issues arising from the free fermion approach. There, the dependence of the coupling constants is plotted for a larger region of distances for selected $\rho_F$. More precisely, the absolute value of the minima, i.e. $- g_{m/\rho_F}(\rho_F)$ for $m\in\mathbbm{N}$  is shown on a double logarithmic plot. From the figure it can be seen, that the long-range decay of the coupling constants is given by
 \begin{equation}
  g_d(\rho_F) \sim \frac{1}{d}.
 \end{equation}
 Concerning the fitting procedure of the couplings to the numerical data it should be mentioned, that the first few distances were left out and that the exponent is slightly less than one because of the finite number of fitting points (When increasing the number of fitted data points, the exponents saturate at one.). This slow decay of the couplings indicated the need for a renormalization procedure, which can be seen from the following argument:\\
 
  \begin{figure}[t]
  \centering
   \Bild[0.95\plotwidth]{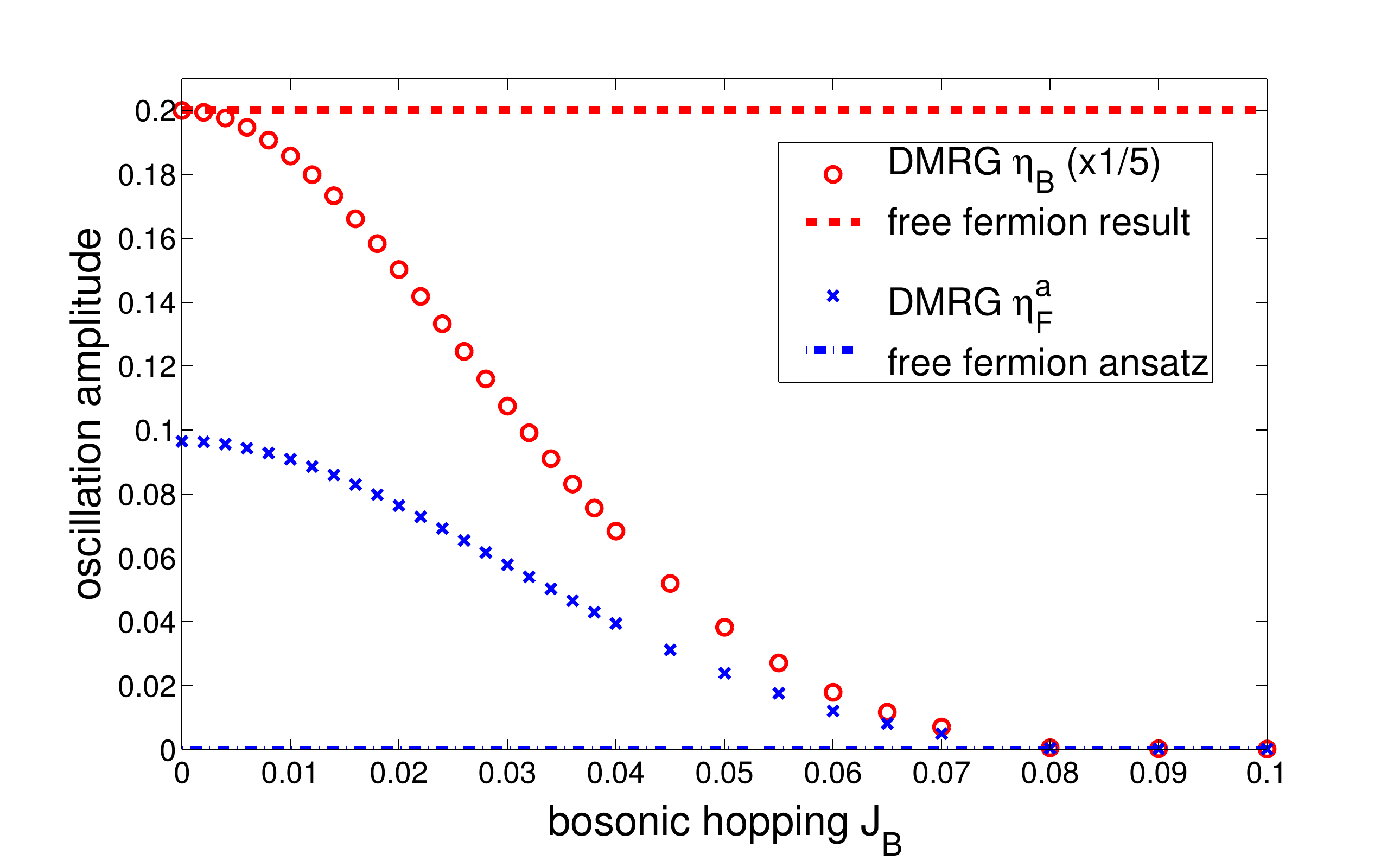}
   \caption{(Color online) Amplitude of the bosonic CDW as a function of the bosonic hopping $J_B$ for $V=1.25$ and $J_F=10$. Since the effective theory predicts a CDW for any hopping $J_B$ (dashed lines), the necessity of a renormalization scheme is evident. Additionally the non-zero amplitude of the fermionic CDW is in strong contrast to the underlying ansatz and another indication of a more involved physics. The numerical data are obtained from DMRG for lattice of $512$ sites and $N_F=N_B=256$.}
   \label{fig:CDWAmplitude}
    \end{figure}
  From the numerical data in figure \ref{fig:PhaseDiagramFastFermions}, we conclude the existence of a CDW phase at double half filling $\rho_F=\rho_B=\frac12$ as also reported in \cite{Pollet2006,Titvinidze2008} for slightly different choice of the system parameters. This CDW phase directly results from the induced interactions and a simple explanation at vanishing bosonic hopping $J_B$ can be found by subsequently adding bosons to the system starting from zero filling up to the CDW filling $\rho_B=\frac12$. The first boson occupies an arbitrary site $j$. A second boson minimizes the energy at site $j\pm2$, since here the density-density interaction is negative. All additional particles will continue occupying all even sites, ending up in the CDW phase at half filling $\rho_B=1/2$. Since the couplings decay as $\frac1d$, the total interaction energy in the thermodynamic limit diverges and since this argument also holds for $J_B>0$, the ground state would always be given by a CDW with full amplitude $\eta_B=1$ for any hopping $J_B$. This result is in strong contrast to the numerical results displayed in figure \ref{fig:PhaseDiagramFastFermions} and more precisely in figure \ref{fig:CDWAmplitude}. The latter one shows the amplitude of the bosonic CDW from figure \ref{fig:PhaseDiagramFastFermions} as a function of the bosonic hopping $J_B$ together with the constant prediction from the argument above. As the amplitude of the CDW quickly drops to zero with in creasing hopping contradicting the previous results, the figure also gives a hint to a solution of this problem. Also shown is the amplitude of a fermionic CDW, i.e., the CDW phase discussed earlier is a double CDW. The existence of a fermionic CDW directly reveals that the initial assumption of free fermions is invalid and the back-action of the bosons to the fermions have to be included, which will be incorporated by the already introduced quantity $\widetilde n_j$. These arguments also hold in the case of a fermionic density $\rho_F\not=\frac12$, with a ground state which has a boson at every $\frac1{\rho_F}$-th site.
   
 Before we move onto a scheme including this back-action for $\rho_F=\frac12$ we consider the properties of the coupling constants in momentum space. This will provide a valuable tool to judge the performance of the upcoming renormalization.\\

\begin{figure}[t]
  \centering
   \Bild{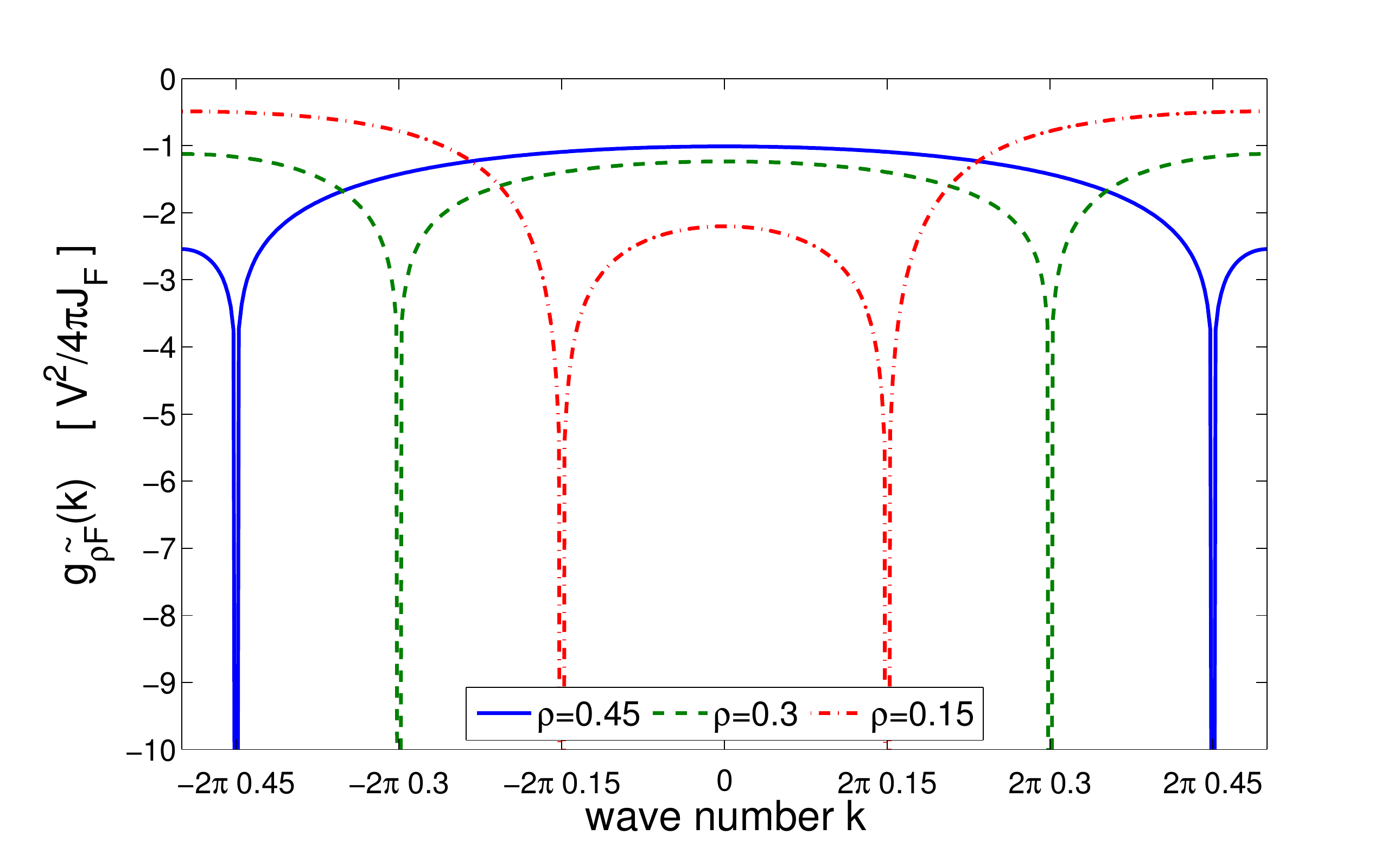}
   \caption{(Color online) Numerical results for the Fourier transform of the couplings from equation (\ref{eq:ResultFourierTrafoFreeFermions}). Shown are the couplings for selected values of $\rho_F$. The divergence at $\pm2\pi\rho_F$ indicating the need for a renormalization of the fermions is very sharp.}
   \label{fig:CouplingsMomentumSpace}
    \end{figure}

The main aspects of the nature of the couplings $g_d(\rho_F)$ can be seen from the Fourier transform of the couplings defined as
\begin{equation}
 \widetilde g_{\rho_F}(k) = \sum_d g_d(\rho_F) e^{i k d}. \label{eq:DefFourierTrafo}
\end{equation}
From the analytic form of the couplings (\ref{eq:CouplingsFreeFermions}), the Fourier transform can be reduced to the Fourier transform of the numerator given by
\begin{multline}
 \sum_{d=-\infty}^\infty \cos d\xi \cos d\xi^\prime e^{i k d} = \\
 \frac\pi 2 \sum_{l=-\infty}^\infty \sum_{C_1,C_2=\pm1} \delta(2\pi l - C_1 \xi - C_2 \xi^\prime-k)
\end{multline}
as proven in section \ref{sec:AppendixFastFourier} in the appendix. By introduction of usual unit step functions $\Theta(x)$, the double integral in Fourier space can be reduced to a single integral
\begin{widetext}
\begin{align}
  \widetilde g_{\rho_F}(k) &= -\frac{V^2}{4\pi J_F}  \sum_{l,C_1,C_2} \int_ {0}^{\rho_F\pi} {\rm d} \xi\ \frac {\Theta(\pi-2\pi l C_2+C_1 C_2 \xi + C_2 k) \Theta(2\pi l C_2-C_1 C_2 \xi - C_2 k-\pi\rho_F) }{ \cos(\xi)-\cos( C_1 \xi+ k)}.\label{eq:ResultFourierTrafoFreeFermions}
   \end{align}
   \end{widetext}
The infinite sum over $l$ turns out to be unproblematic since the $\Theta$-functions strongly limit the valid range of $l$ and naturally the couplings are $2\pi$-periodic. For $k=0$, above integral expression is undefined, where a proper treatment of the limit $k\to0$ gives $\widetilde g_{\frac12}(0)=-\frac{V^2}{4\pi J_F}$. Figure \ref{fig:CouplingsMomentumSpace} shows the couplings in momentum space as a function of the momentum $k$ for different densities $\rho_F$. From the figure, a divergence for $k=\pm2\pi\rho_F$ may be seen. This van Hove-singularity \cite{vanHove1953}, also reported for instance in \cite{Buechler2003,Buechler2004} is directly connected to the earlier discussed divergence of the energy.

   \section{Renormalization of the fermionic system and the effective Hamiltonian}\label{chap:RenormalizationFermions}\label{chap:RenormalizedHamiltonian}
   The previous section proved that the initial ansatz, assuming a full decoupling of the fermions leads to an unphysical behavior. This problem is overcome in the present section. As already stated above, the induced interaction on the free fermion level drives the bosons into a CDW state. This CDW state now, in turn, acts as an external potential to the fermionic subsystem. Thus, introducing this back-action, the fermionic correlators have to be calculated with respect to fermions in an alternating potential, introduced into the Hamiltonian by the bosonic mean-field amplitude $\widetilde n_j$ in equation (\ref{eq:FermionicHamiltonian}). In the following we restrict ourselves to the most interesting case $\rho_F=1/2$, where a generalization to other situations with $\rho_F=1/m$ with $m\in\mathbbm{N}$ will be provided elsewhere.

Treating the bosons as being in a CDW state, the mean-field amplitude ansatz \footnote{A similar ansatz is used in \cite{Pazy2005} to study the influence of the wavelength of the bosonic CDW on the fermionic system.} is given by
\begin{align}
 \widetilde n_j&=\rho_B\left[1+\eta_B (-1)^j\right]\label{eq:AnsatzBosonicCDW}\\
 &=\rho_B(1-\eta_B)+2\rho_B\eta_B\   \delta( \sin (\pi \frac j2)),
\end{align}
where the latter form serves as a simplification in the following calculation. Here we introduced the amplitude of the bosonic CDW $\eta_B$ as a free parameter. Figure \ref{fig:CDWAmplitude} already showed that this amplitude drops to zero for increasing hopping $J_B$.\\

The main task in resolving the divergence is to calculate the fermionic cumulants used in the effective bosonic Hamiltonian (\ref{eq:effectiveBHMFull}), i.e., free fermions in an alternating potential, given by
\begin{multline}
 \H F = -J_F\sum_j\left( \hat c^\dagger_j\hat c_{j+1}+\hat c^\dagger_{j+1}\hat c_j\right)\label{eq:FermionicHamiltonianRenormalized}\\
 + V2\rho_B\eta_B\sum_j  \delta( \sin \pi \frac j2)\nf_j. 
 \end{multline}
 In this Hamiltonian, a global energy shift $V\rho_B (1-\eta_B)\rho_F$  from the potential is left out and a solution can be found straightforwardly. Although the solution is easy by means of a canonical transformation \cite{Rousseau2006,Lieb1961}, the resulting expressions are rather involved and the quantities needed are hard to express. Here we employ a Green's function approach, extracting all needed quantities for the full calculation of the bosonic Hamiltonian for double half filling.

\subsection{General framework and initial definitions}\label{sec:DysonEquation}
In order to calculate the second order cumulant $\langle\langle\nf_j(T)\nf_{j+d}(0)\rangle\rangle_{\rm F}$ with respect to the ground state of the fermionic Hamiltonian (\ref{eq:FermionicHamiltonianRenormalized}) we make use of the Green's function technique \cite{Mahan2000}. The second order cumulant factorizes by use of Wick's theorem \cite{Mahan2000,Wick1950} into a product of advanced and retarded Green's functions
\begin{align}
 \langle\langle\nf_j(T)\nf_{j+d}(0)\rangle\rangle_{\rm F} &=\mathcal G^{(+)}_{j,j+d}(t+T,t)\   \mathcal G^{(-)}_{j,j+d}(t+T,t)\notag\\
 &\hspace{-2cm}=  \left\langle \hat c^\dagger_j(t+T)\hat c_{j+d}(t)\right\rangle\left \langle \hat c_{j}(t+T)\hat c^\dagger_{j+d}(t) \right\rangle.\label{eq:DensityDensitySplitted}
\end{align}
Here we used $T>0$ and the definition of the Green's functions
\begin{equation}
\begin{split}
   \mathcal G^{(+)}_{j,j+d}(t+T,t)&=<\mathcal T\ \hat c^\dagger_j(t+T)\ \hat c_{j+d}(t)>,\\
   \mathcal G^{(-)}_{j,j+d}(t+T,t)&= <\mathcal T\ \hat c_j(t+T)\ \hat c^\dagger_{j+d}(t)>.
   \end{split}\label{eq:DefinitionGreensFunctionRealSpace}
\end{equation} 
To find a solution of the problem it is more convenient to switch to momentum space. Using the Fourier transformation (\ref{eq:FourierTrafo}), Hamiltonian (\ref{eq:FermionicHamiltonianRenormalized}) gives
\begin{multline}
  \H F =-2J_F\sum_{k=-L/2}^{L/2-1} \cos(2\pi\frac kL)\, \fd_k \f_k\\
  +V\eta_B\rho_B  \sum_{k=-L/2}^{L/2-1}\sum_{\alpha=\pm1} \, \fd_{k+\frac L2\alpha} \f_{k}\label{eq:FermionicHamiltonianRenormalizedFourier}
\end{multline}
apart from a constant energy shift $V \eta_B\rho_B \rho_F$ which is neglected. Here it should be mentioned that the summation over $\alpha$ only includes those terms which fulfill $|k|<\frac L2$. To denote Green's functions in momentum space change indices as $j\to k$ and $j+d\to k^\prime$.\\

Due to the perturbation of the ground state from the potential $V$, we first calculate the Green's functions (\ref{eq:DefinitionGreensFunctionRealSpace})  for the unperturbed system, i.e., the ground state of Hamiltonian (\ref{eq:FermionicHamiltonianRenormalized}) for $\eta_B=0$. A straightforward calculation gives for the Green's functions
\begin{align}
 \mathcal G^{(0+)}_{k,k^\prime}(t,t^\prime)  &=\Theta(t-t^\prime)\Theta(\epsilon_F-\epsilon_k)e^{i\epsilon_k(t-t^\prime)}\delta_{k,k^\prime}\notag\\
 &\hspace{1cm}-\Theta(t^\prime-t)\Theta(\epsilon_k-\epsilon_F)e^{i\epsilon_k(t-t^\prime)}\delta_{k,k^\prime}\notag\\
  \mathcal G^{(0-)}_{k,k^\prime}(t,t^\prime) &=\Theta(t-t^\prime)\Theta(\epsilon_k-\epsilon_F)e^{i\epsilon_k(t^\prime-t)}\delta_{k,k^\prime} \label{eq:UnperturbedGreensFunctionMomentumSpace}\\
  &\hspace{1cm}-\Theta(t^\prime-t)\Theta(\epsilon_F-\epsilon_k)e^{i\epsilon_k(t^\prime-t)}\delta_{k,k^\prime}\notag
  \end{align}
in the time domain and 
\begin{equation}
\mathcal G^{(0\pm)}_{k,k^\prime}(\omega)=\pm\delta_{k,k^\prime}\frac{i}{\sqrt{2\pi}} \frac{1}{\epsilon_k\mp\omega\oplus i\delta} \label{eq:GreensFunctionFreeFermions}
\end{equation}
in the frequency domain. Here, the frequency domain is defined by the (time) Fourier transformation
\begin{equation}
  \mathcal G^{(0\pm)}_{k,k^\prime}(\omega) =\frac{1}{\sqrt{2\pi}}\int_{-\infty}^\infty {\rm d}T\, \mathcal G^{(0\pm)}_{k,k^\prime}(t+T,t)\  e^{-i\omega T}\ e^{\pm\delta T}.
\end{equation}
The last term in the integral kernel is introduced to assure convergence and will be properly removed later on. In equation \eqref{eq:UnperturbedGreensFunctionMomentumSpace} we introduced the dispersion relation $\epsilon_k=-2J_F\cos(2\pi\frac kL)$ of the free particle and 
\begin{equation}
 \oplus\ =   \begin{cases}
  +  \hspace{1cm}&k\in\mathcal K_F\\
  -  \hspace{1cm}&k\not\in\mathcal K_F
  \end{cases}
\end{equation}
distinguishes between momentum modes within the Fermi sphere and those outside.\\

Following the technical details presented in \cite{Mahan2000}, we immediately arrive at a Dyson equation for the Green's function since the induced potential is only quadratic in the fermionic operators. This gives

\begin{align}
 \mathcal G^{(+)}_{k,k^\prime}(\omega)  &= \mathcal G^{(0+)}_{k,k^\prime}(\omega)\\
 &\hspace{0.1cm}+\frac{i}{\hbar} \sqrt{2\pi}V\eta_B\rho_B\ \mathcal G^{(0+)}_{k,k}(\omega)   \sum_{\alpha=\pm1}  \mathcal G^{(+)}_{k+\frac L2\alpha,k^\prime}(\omega)\label{eq:DysonGreensAdvanced}\notag
 \intertext{for the advanced Green's function and the retarded Green's function is given by}
 \mathcal G^{(-)}_{k,k^\prime}(\omega)  &= \mathcal G^{(0-)}_{k,k^\prime}(\omega)\\
 &\hspace{0.1cm}+\frac{i}{\hbar} \sqrt{2\pi}V\eta_B\rho_B \ \mathcal G^{(0-)}_{k^{\prime},k^\prime}(\omega) \sum_{\alpha=\pm1} \mathcal G^{(-)}_{k,k^{\prime}+\frac L2\alpha}(\omega).\notag
\end{align}
These equations allow for an algebraic solution.

\subsection{Solution of the Dyson equations}\label{sec:SolutionDysonEquation}
For the solution we only discuss the case of the advanced Green's function, the situation for the retarded one is exactly the same. Considering
\begin{align}
 \mathcal G^{(+)}_{k\pm\frac{L}{2},k^\prime}(\omega)  &= \mathcal G^{(0+)}_{k\pm\frac{L}{2},k^\prime}(\omega)\\
 &\hspace{0.1cm}+\frac{i}{\hbar} \sqrt{2\pi}V\eta_B\rho_B \, \mathcal G^{(0+)}_{k\pm\frac{L}{2},k\pm\frac{L}{2}}(\omega) \mathcal G^{(+)}_{k,k^\prime}(\omega),\notag
\end{align}
the contributions in the latter part of equation (\ref{eq:DysonGreensAdvanced}), reinserting them into (\ref{eq:DysonGreensAdvanced}) and solving for $\mathcal G^{(+)}_{k,k^\prime}(\omega)$, the solution of the Green's functions in terms of the unperturbed ones is given by
\begin{widetext}
\begin{equation}
   \mathcal G^{(+)}_{k,k^\prime}(\omega) =\frac{\mathcal G^{(0+)}_{k,k^\prime}(\omega)+\frac{i}{\hbar} \sqrt{2\pi}V\eta_B\rho_B
 \, \mathcal G^{(0+)}_{k,k}(\omega)\left[ \mathcal G^{(0+)}_{k+\frac{L}{2},k^\prime}(\omega) +  \mathcal G^{(0+)}_{k-\frac{L}{2},k^\prime}(\omega) \right]}{1+\frac{ 2\pi V^2\eta_B^2\rho_B^2}{\hbar^2}\mathcal G^{(0+)}_{k,k}(\omega)\left[  \mathcal G^{(0+)}_{k+\frac{L}{2},k+\frac{L}{2}}(\omega)+ \mathcal G^{(0+)}_{k-\frac{L}{2},k-\frac{L}{2}}(\omega)\right]}.
\end{equation}
\end{widetext}
Here it should be mentioned that the contribution from $\mathcal G^{(+)}_{k\pm L ,k^\prime}(\omega)$ vanish since the momentum modes are limited to the first Brillouin zone $k\in[-\frac L2,\frac L2]$ with $k\pm L \not \in [-\frac L2,\frac L2]$.

From \eqref{eq:GreensFunctionFreeFermions} we note that $\mathcal G^{(0\pm)}_{k,k^\prime}(\omega)\sim\delta_{kk^\prime}$ and therefore only the terms $ \mathcal G^{(+)}_{k,k}(\omega)$, $ \mathcal G^{(+)}_{k,k\pm\frac L2}(\omega)$ and $ \mathcal G^{(+)}_{k\pm\frac L2,k}(\omega)$ of the full Green's functions are non-zero. Applying the same procedure to $ \mathcal G^{(-)}_{k,k^\prime}(\omega)$ gives a similar expression. The final form for the Green's functions in momentum space and time domain is found by using the precise form of  $\mathcal G^{(0\pm)}_{k,k^\prime}$ from (\ref{eq:GreensFunctionFreeFermions}) and simplifying the resulting expressions, giving
\begin{equation}
 \mathcal G^{(\pm)}_{k,k}(\omega) =\pm\frac{i}{\sqrt{2\pi}}\frac{\epsilon_k\pm\omega\oplus i\delta}{(\epsilon_k\mp\omega\oplus i\delta)(\epsilon_k\pm\omega\oplus i\delta)+\frac{V^2\eta_B^2\rho_B^2}{\hbar^2}}\label{eq:GreensSolutionMomentumFrequencyEqual}
\end{equation}
and
\begin{align}
 \mathcal G^{(\pm)}_{k,k\pm\frac{L}{2}}(\omega) &=\mathcal G^{(\pm)}_{k\pm\frac{L}{2},k}(\omega)\label{eq:GreensSolutionMomentumFrequencyShifted}\\
 &=\frac{i \frac{V\eta_B\rho_B} {\sqrt{2\pi}\hbar}}{(\epsilon_k\mp\omega\oplus i\delta)(\epsilon_k\pm\omega\oplus i\delta)+\frac{V^2\eta_B^2\rho_B^2}{\hbar^2}} .\notag
\end{align}
 Here, again $\oplus$ distinguishes between momentum modes $k$ within or outside the Fermi sphere $\mathcal K_F$.\\
 
 This is the main result of this section. It has to be completed by transforming back to the time domain which can be found in appendix \ref{sec:AppendixFourierTrafoGreens}. After the Fourier transformation, introducing the renormalized dispersion relation
 \begin{equation}
  \bar\epsilon_k=\sqrt{\epsilon_k^2+\frac{V^2\eta_B^2\rho_B^2}{\hbar^2}}, \label{eq:DispersionRenomalized}
 \end{equation}
 the Green's functions in momentum space and time domain are found to read
 \begin{gather}
  \mathcal G^{(\pm)}_{k,k}(t+T,t) = \frac{1 }{2}e^{-i \bar\epsilon_k T}\left(1\mp\frac{\epsilon_k}{\bar\epsilon_k}\right)\label{eq:GreensSolutionMomentumTimeEqual}\\
 \mathcal G^{(\pm)}_{k\pm\frac{L}{2},k}(t+T,t) =-\frac{V\eta_B\rho_B}{2\hbar}\frac{1}{\bar\epsilon_k}e^{-i \bar\epsilon_k T}.\label{eq:GreensSolutionMomentumTimeShifted}
 \end{gather}
These expressions allow to calculate the density-density correlations for the fermionic Hamiltonian according to \eqref{eq:DensityDensitySplitted}.
 
 \subsection{Green's function in real space and expectation values}
 Going back from momentum to real space allows for a direct calculation of the needed cumulant. Again we restrict ourselves on the calculation of $\mathcal G^{(+)}_{j,j+d}(t+T,t)$, since the calculation for $\mathcal G^{(-)}_{j,j+d}(t+T,t)$ is similar. Following the definitions \eqref{eq:DefinitionGreensFunctionRealSpace} of the Green's functions and (\ref{eq:FourierTrafo}) of the Fourier transformation, the real space Green's functions are connected to the momentum space Green's functions by
 \begin{align}
  \mathcal G^{(+)}_{j,j+d}(t+T,t)&= \frac{1}{L}\sum_{k_1,k_2=-L/2}^{L/2-1}  \\
  &\hspace{-0.3cm}e^{-2\pi i\frac{(k_1-k_2)}{L}j}e^{-2\pi i\frac{k_1}{L}d}{\mathcal G^{(+)}_{k_2,k_1}(t+T,t) }.\notag
\end{align}
From the previous discussions we know, that only certain Green's functions in momentum space are non-zero. This is incorporated by including $ \delta_{k_1k_2}+\delta_{k_1+\frac{L}{2}k_2}+\delta_{k_1-\frac{L}{2}k_2}$ to the summation, picking out the non-zero elements. After performing the summation over $k_2$, a slight restructuring of the exponentials and an application of the symmetry of the Green's functions (\ref{eq:GreensSolutionMomentumFrequencyShifted}), the final result in the thermodynamic limit is found to be

 \begin{widetext}
\begin{equation}
  \mathcal G^{(\pm)}_{jj+d}(t+T,t)=\frac{1}{2\pi}\int_{0}^{\pi}{\rm d}\xi\,\cos( d\xi)\ e^{- iT2J_F\sqrt{\cos^2(\xi)+a^2}}\left(1\pm\frac{\cos(\xi)}{\sqrt{\cos^2( \xi)+a^2}}\right)-(-1)^j\frac{a}{2\pi}\int_{0}^{\pi}{\rm d}\xi\,\cos(d\xi)\ \frac{e^{- iT2J_F\sqrt{\cos^2(\xi)+a^2}}}{\sqrt{\cos^2(\xi)+a^2}}.
\label{eq:GreensFunctionRenormalizedFermions}
\end{equation}
\end{widetext}
Here we introduced  an amplitude factor $a=\frac{V\eta_B\rho_B}{2\hbar J_F}$. Note that the integration cannot be carried out explicitly for arbitrary distance $d$.

The calculation of the Green's functions does not only allow to calculate the density-density correlator in equation (\ref{eq:CouplingsGeneral}) but also gives a prediction of the behavior of the fermionic system, as long as the bosonic CDW amplitude $\eta_B$ is known. Here we first verify the analytic expression of the Green's function in the fermionic problem itself, i.e., all numerical data shown are calculated for the Hamiltonian (\ref{eq:FermionicHamiltonianRenormalized}).\\

\begin{figure}[t]  
  \centering
   \Bild{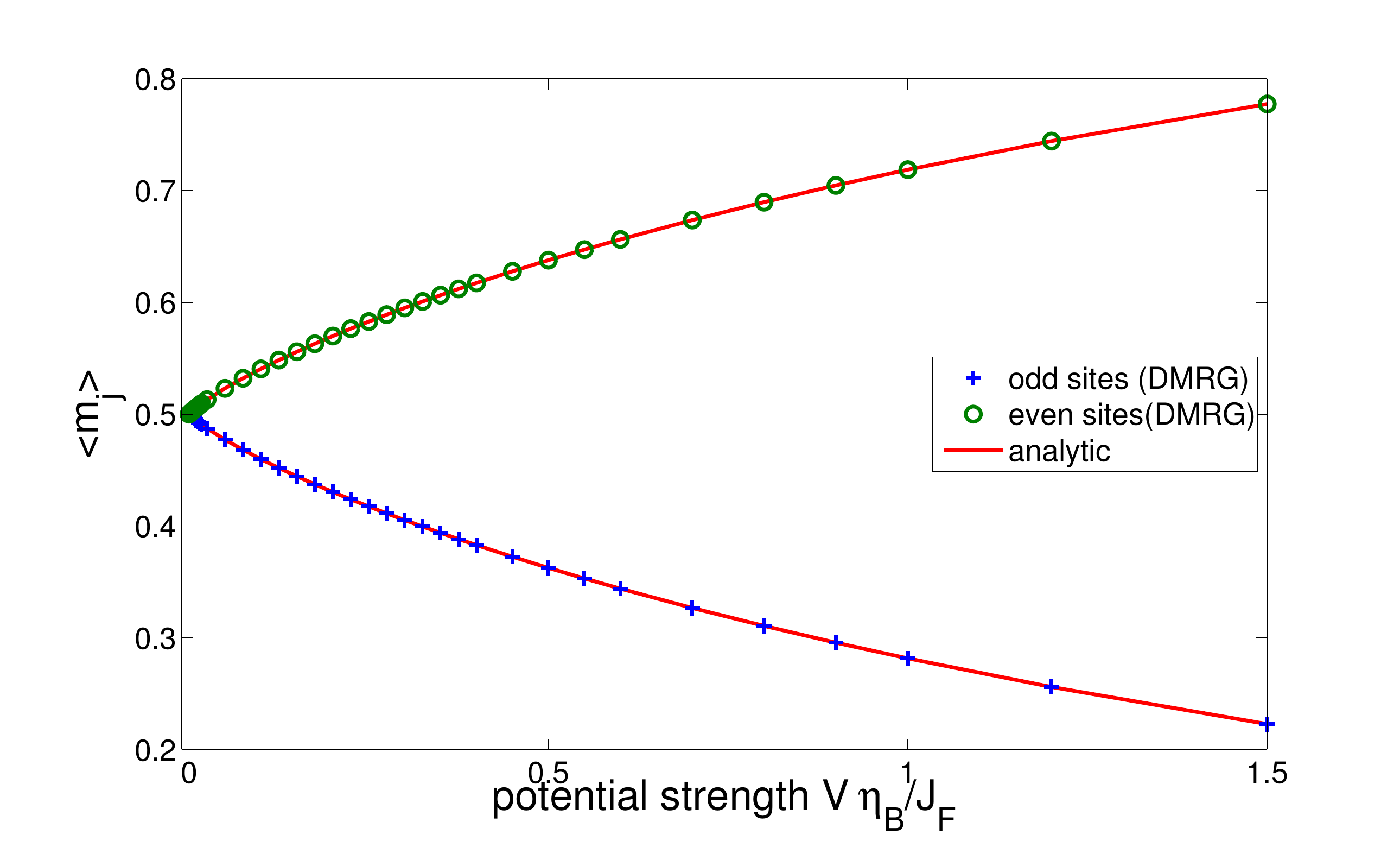}
   \caption{(Color online) Expectation value of the fermionic density operator for even and odd sites for the ground state of  the effective renormalized fermionic Hamiltonian (\ref{eq:FermionicHamiltonianRenormalized}). Points are the numerical results from DMRG calculations with $300$ sites and $J_F=10$. Solid lines are the analytic results from equation (\ref{eq:FermionicDensity}).  Shown are the numerical results for     $\langle\nf_{150}\rangle_{\rm F}$ and $ \langle\nf_{151}\rangle_{\rm F} $. }
    \label{fig:FermionicCDWAmplitude}
    \end{figure}

\textbf{Local density:} The expression for the Green's functions gives an (analytic) prediction of the fermionic density in the alternating potential. Using $\langle\nf_j\rangle_{\rm F} =  \mathcal G^{(+)}_{j,j+0}(t+0,t)$, the fermionic density evaluates analytically as
\begin{equation}
  \langle\nf_j\rangle_{\rm F} =\frac{1}{2}- (-1)^j\frac{a}{\pi\sqrt{1+a^2}}\ K\left[\frac{1}{1+a^2} \right].
\end{equation}
The first important result from the renormalization procedure therefore is
\begin{equation}
\langle\nf_j\rangle_{\rm F}=\frac12\left[1-\eta_F^a (-1)^j\right],\label{eq:FermionicDensity}
\end{equation}
where $\eta_F^a=\frac{2a}{\pi\sqrt{1+a^2}}K\left[\frac{1}{1+a^2} \right]$ and $K[x]$ is the complete elliptic integral of the first kind \cite{Abramowitz1964}. This means, the renormalization procedure results in the prediction of a fermionic CDW with some amplitude $\eta_F^a$ which is in agreement with the numerical results from figure \ref{fig:CDWAmplitude}. Figure \ref{fig:FermionicCDWAmplitude} shows numerical calculation of the amplitude of the fermionic CDW from DMRG calculations for the Hamiltonian (\ref{eq:FermionicHamiltonianRenormalized}) as a function of the potential strength $V\eta_B$ along with the analytic results. 

Another feature of \eqref{eq:FermionicDensity} which will be important for the later discussion of the full BFHM is the minus sign in front of the site dependent part. This is a direct consequence of the alternating boson potential ansatz. Since the interaction $V$ is chosen positive, i.e., repulsion between bosons and fermions, it is expected that the phase of the bosonic and fermionic density wave is shifted by $\pi$ compared to each other. For the case of attractive interaction, both density waves are in phase. This is in full agreement with the numerical results presented in the discussion of the results for the full BFHM in chapter \ref{chap:ResultsFastFermionsFullModel}. In the limit $a\to 0$, corresponding to the free fermion case the result for the density reduces to the result for free fermions at half filling, i.e., $ \langle\nf_j\rangle_{\rm F}=\frac12$.

  \begin{figure}[t]  
  \centering
   \Bild{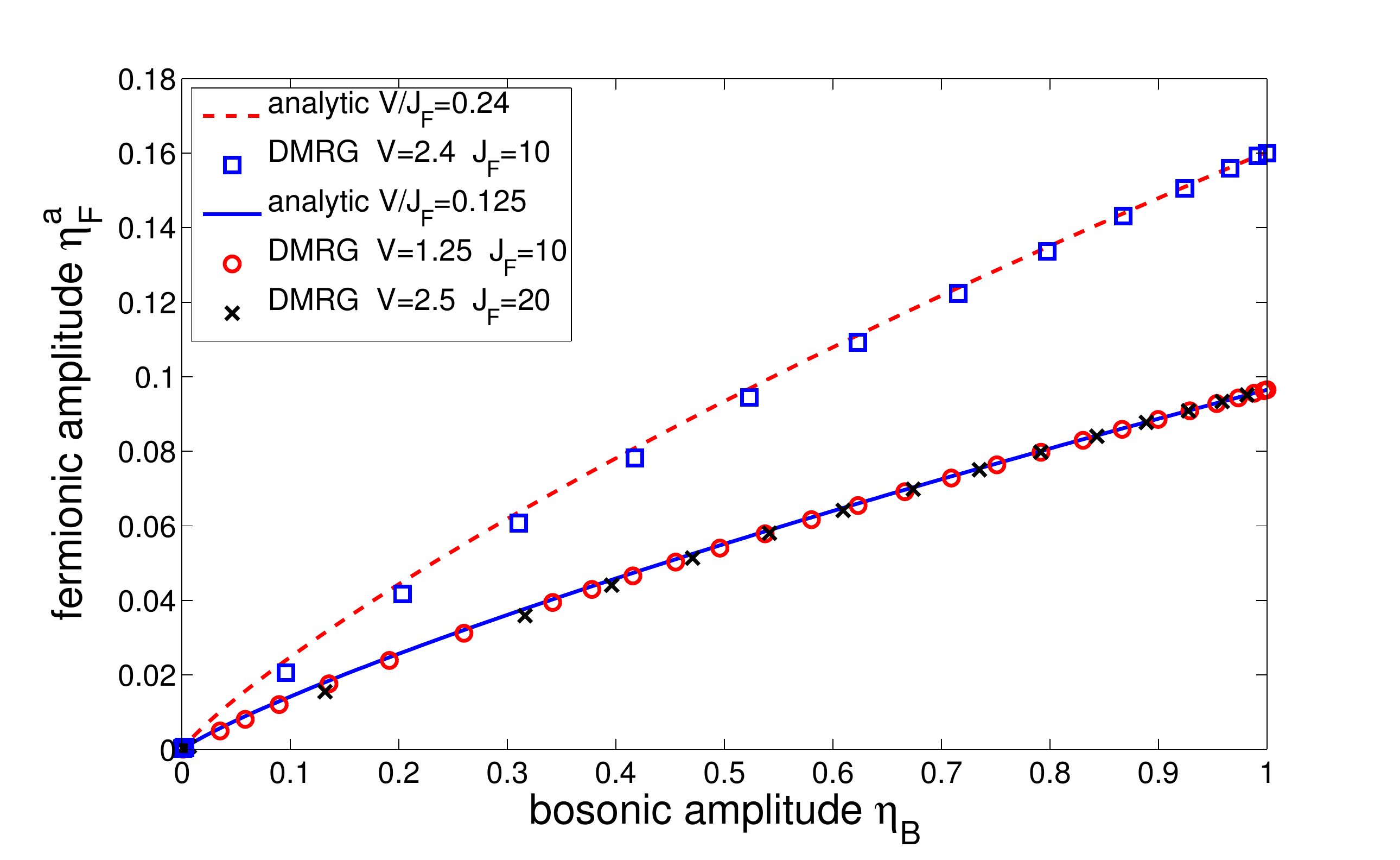}
   \caption{(Color online) Amplitude $\eta_F^a$ of the fermionic CDW versus the bosonic amplitude $\eta_B$ for different numerical data. Shown are the numerical results (data points) presented in figure \ref{fig:CDWAmplitude} and for $V=2.4$ and $J_F=10$ obtained from the full BFHM. The solid lines are the analytic results for $\eta_F^a$. }
    \label{fig:EtaBversusEtaF}
    \end{figure}
  
  Figure \ref{fig:EtaBversusEtaF} furthermore shows the numerical results for the amplitudes $\eta_F^a$ as a function of $\eta_B$ from figure \ref{fig:CDWAmplitude} as well as the analytic prediction according to equation (\ref{eq:FermionicDensity}). To remember, the numerical data comes from the full BFHM, proving the chosen approach to be valid already in the prediction of the fermionic quantities.\\

\begin{figure}[t]  
  \centering
  \Bild[\plotwidth]{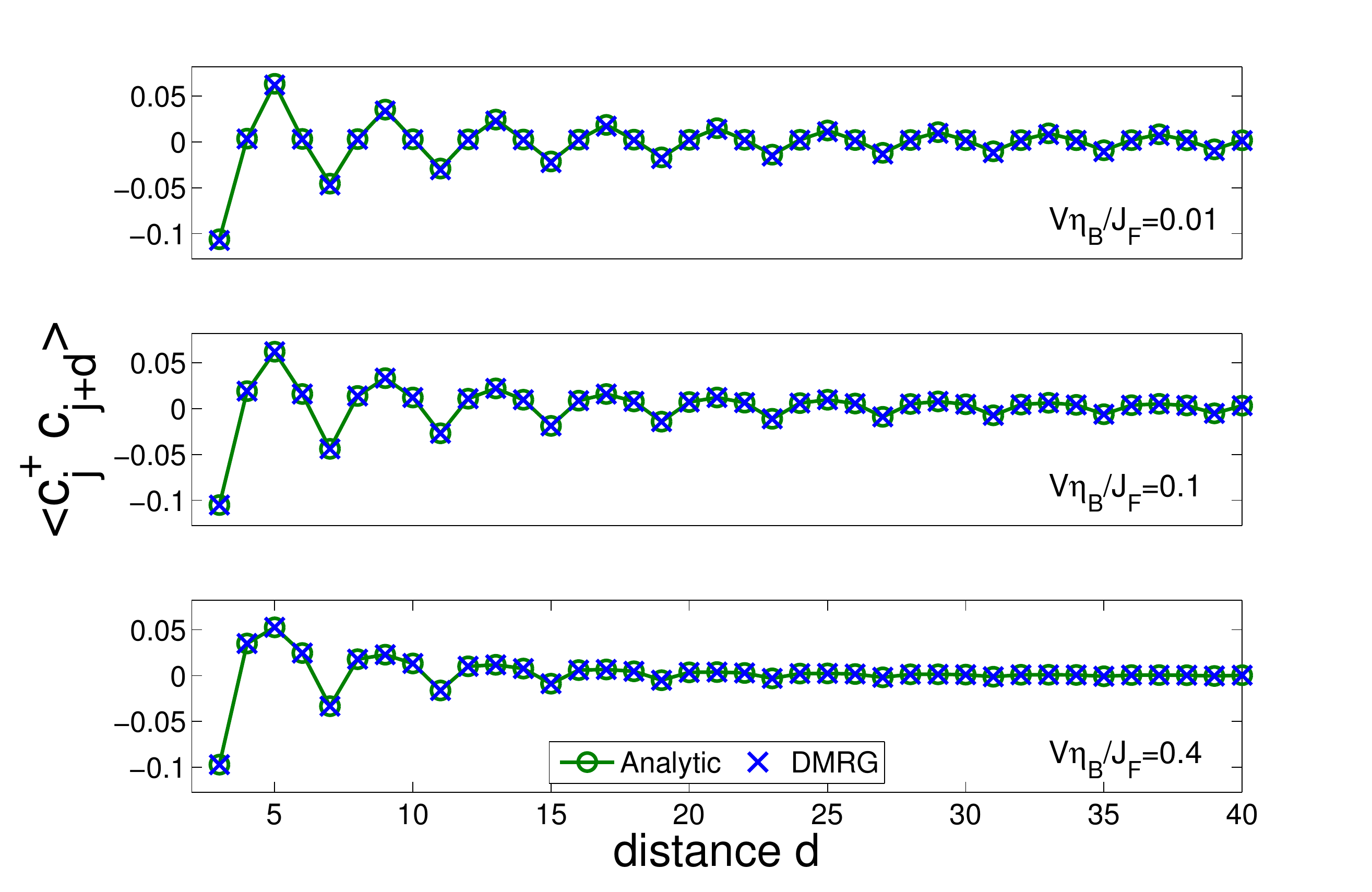}
    \caption{(Color online) Distance dependence of the first-order correlations $ <\hat c^\dagger_j\ \hat c_{j+d}> $  for three different interactions $V\eta_B$ calculated from the fermion model (\ref{eq:FermionicHamiltonianRenormalized}). Solid lines are the theoretical results from a numerical integration of \eqref{eq:GreensFunctionRenormalizedFermions}. The points are the numerical results from the data used in figure \ref{fig:FermionicCDWAmplitude}. }
    \label{fig:FermionicCorrelationsa}
    \end{figure}

\begin{figure}[t]  
  \centering
\Bild[\plotwidth]{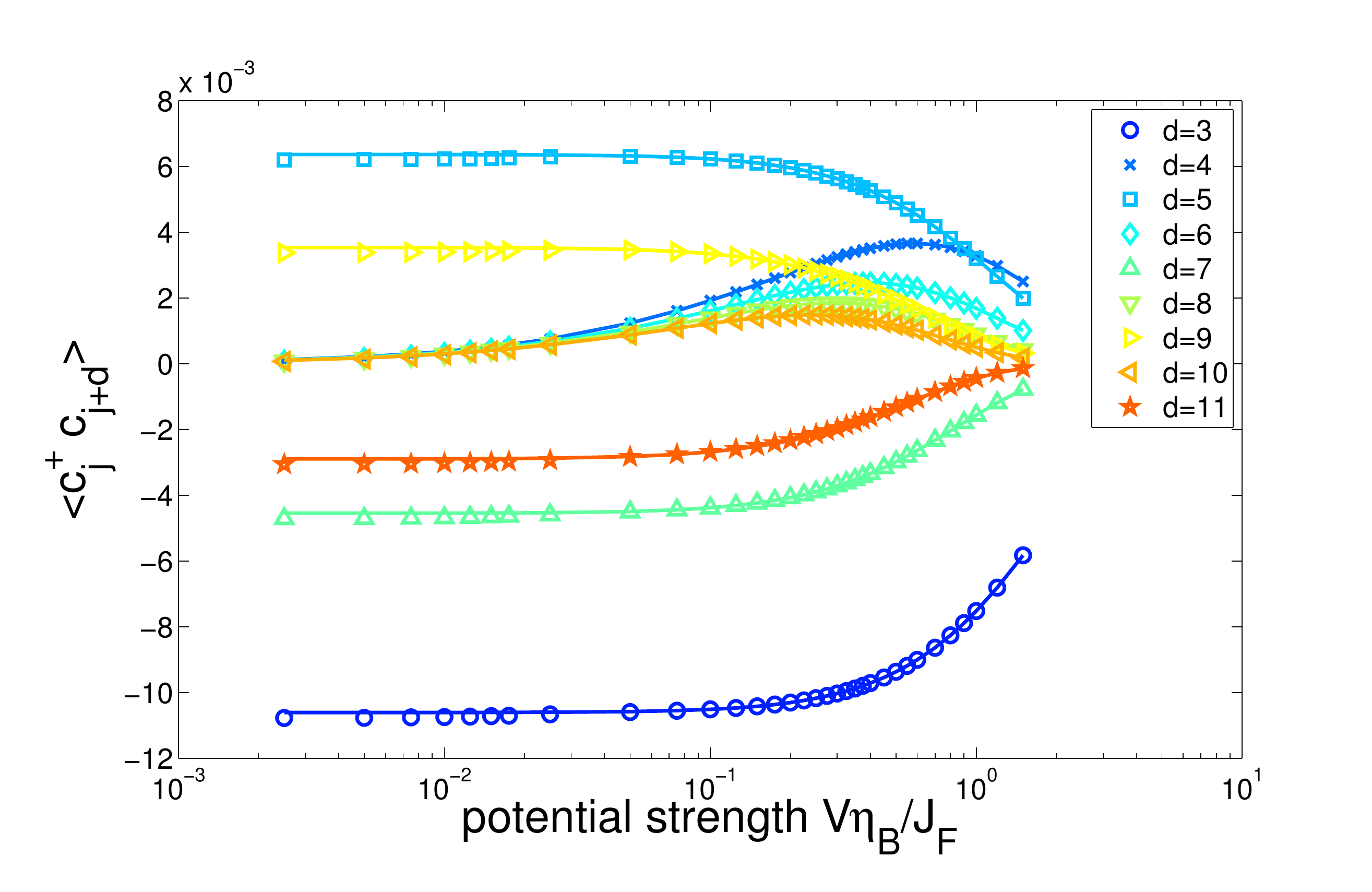}
    \caption{(Color online) Dependence of the first-order correlations $ <\hat c^\dagger_j\ \hat c_{j+d}> $ for a wide range of interactions $V\eta_B$ for selected distances $d$ calculated from the fermion model (\ref{eq:FermionicHamiltonianRenormalized}). Solid lines are the theoretical results from a numerical integration of \eqref{eq:GreensFunctionRenormalizedFermions}. The points are the numerical results from the data used in figure \ref{fig:FermionicCDWAmplitude}.}
    \label{fig:FermionicCorrelationsb}
    \end{figure}

  \textbf{First-order correlations:} Figures \ref{fig:FermionicCorrelationsa} and \ref{fig:FermionicCorrelationsb} show numerical results for the  first-order correlations $  \left\langle\hat c^\dagger_j\ \hat c_{j+d}\right\rangle =   \mathcal G^{(+)}_{j,j+d}(t+0,t)$ compared to the analytic results. Unfortunately, the integral expression for the Green's function cannot be evaluated analytically for arbitrary distance $d$, making a numerical integration necessary.  The perfect agreement proves the validity of the obtained expression for the Green's function.\\
  
  
  \textbf{Density-density correlations:}  Finally we calculate the density-density correlations used in the expression for the coupling constants (\ref{eq:CouplingsGeneral}) with the renormalized fermionic model. Having a closer look at the result for the Green's function (\ref{eq:GreensFunctionRenormalizedFermions}) it can be seen that both Green's functions are of the general form $\mathcal G^{(\pm)}_{j,j+d}(t+T,t) = A_\pm - a B$. In equation (\ref{eq:DensityDensitySplitted}) we already noted that the density-density cumulant splits up into the product of advanced and retarded Green's function, which may be written as
  \begin{equation}
    \langle\langle\nf_j(T)\nf_{j+d}(0)\rangle\rangle_{\rm F} = A_+ A_- - a(A_++A_-) + a^2 B^2.\label{eq:SimplifikationDensityDensity}
  \end{equation}
 From the definition of the coupling constants (\ref{eq:CouplingsGeneral}) we can see, that they are proportional to $V^2$, since they are a second order correction in the effective Hamiltonian (\ref{eq:effectiveBHMFull}). This means, that in order $V^2$, only the first term in \eqref{eq:SimplifikationDensityDensity} is relevant.\\
   
Following this argument, the renormalized form of the density-density cumulant reads
  \begin{align}
  \langle\langle \hat m_j(t+T) \hat m_{j+d}(t)\rangle\rangle &=\frac{1}{4\pi^2}\int_{0}^{\pi}\int_{0}^{\pi}{\rm d}\xi{\rm d}\xi^\prime\,\cos(d\xi )\,\cos(d\xi^\prime )\notag\\
 & \hspace{-1.5cm} \times e^{-iT2J_F\left( \sqrt{\cos^2(\xi)+a^2}+\sqrt{\cos^2(\xi^\prime)+a^2} \right)}\label{eq:CumulantRenormalizedFermions}\\
 &\hspace{-3cm}\times\left(1+ \frac{\cos(\xi)}{\sqrt{\cos^2( \xi)+a^2}}\right)\left(1-\frac{\cos(\xi^\prime)}{\sqrt{\cos^2( \xi^\prime)+a^2}}\right) .\notag
   \end{align}
  This is the main result from the renormalization procedure. Comparing the renormalized result to that of free fermions (at $\rho_F=\frac12$) in equation (\ref{eq:CumulantFreeFermions}) one can see, that the corresponding limit $a\to0$ gives the same result as equation (\ref{eq:CumulantFreeFermions}). Note, that the last line in (\ref{eq:CumulantRenormalizedFermions}) serves as a cutoff function which constrains the  integration limits to the free fermion values in the limit $a\to0$.

\subsection{Discussion of the renormalized couplings}

Applying the time integration from \eqref{eq:CouplingsGeneral} as done in the previous section using the Riemann-Lebesgue lemma, the renormalized couplings  $g_d(a)$ at half fermionic filling $\rho_F=1/2$ can be found to be

\begin{figure}[t]
  \centering
   \Bild{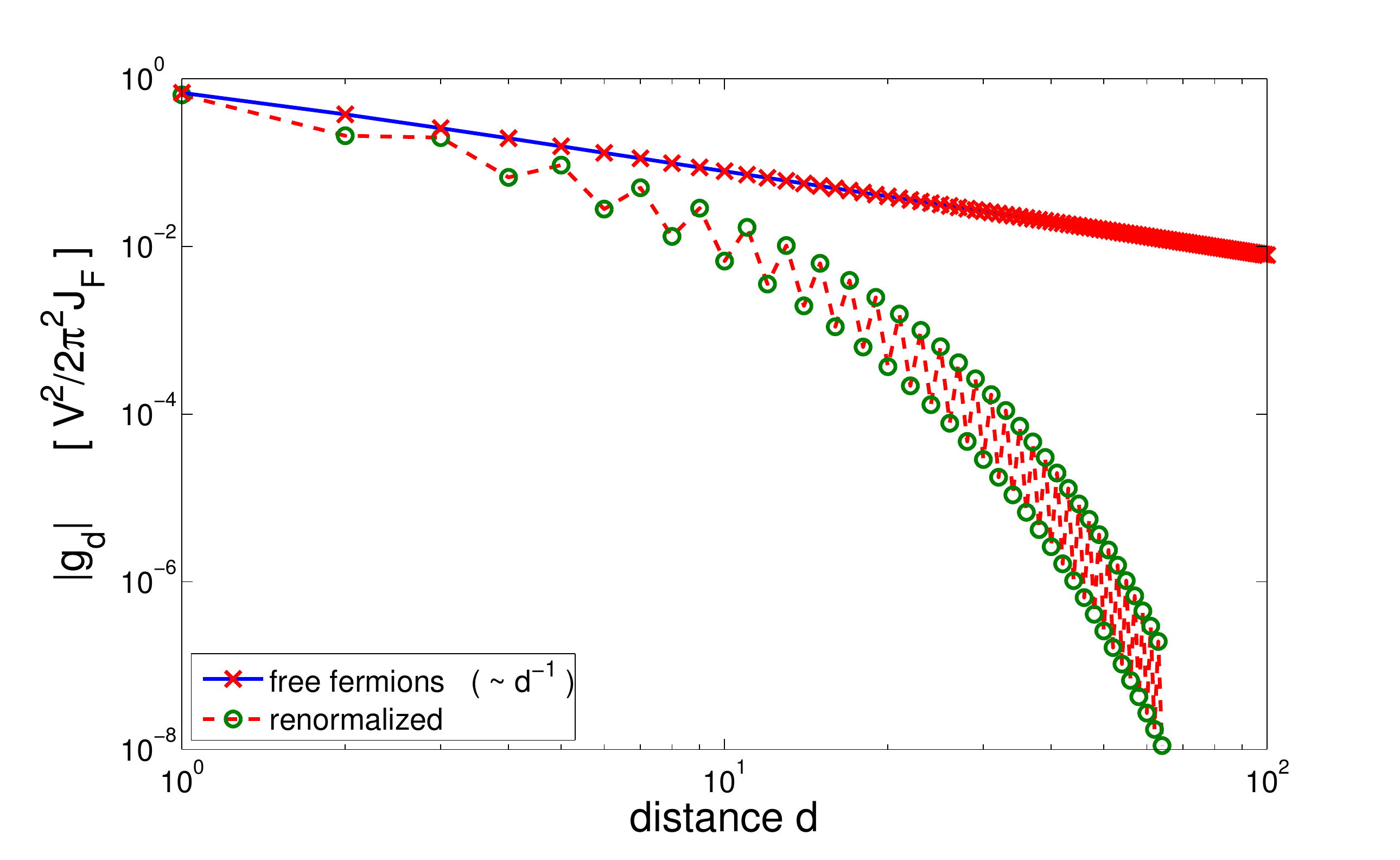}
   \caption{(Color online) Comparison of the couplings for the free fermion case ($a=0$) and the renormalized couplings for $a=0.1$. The free fermion couplings decay as $\frac 1d$, whereas the renormalized couplings decay much faster, preventing the divergence of the energy for the ground state.}
   \label{fig:ComparisonCouplings}
    \end{figure}

\begin{widetext}
\begin{equation}
  g_d(a) = -\frac{V^2}{8 \pi^2J_F }\int_{0}^{\pi}\int_{0}^{\pi}{\rm d}\xi{\rm d}\xi^\prime\,  \frac{\cos(d\xi)\,\cos(d\xi^\prime)}{ \sqrt{\cos^2(\xi)+a^2}+\sqrt{\cos^2(\xi^\prime)+a^2}}\left(1+ \frac{\cos(\xi)}{\sqrt{\cos^2( \xi)+a^2}}\right)\left(1-\frac{\cos(\xi^\prime)}{\sqrt{\cos^2( \xi^\prime)+a^2}}\right).\label{eq:CouplingsRenormalized}
\end{equation}
\end{widetext}  
Since we restricted ourselves to the case of half filling for the fermions, the additional argument $\rho_F$ is dropped
  here but the dependence of the renormalized couplings on the amplitude factor $a$ is explicitly written. As done in the case of free fermions, we shortly discuss the properties of the renormalized couplings. Figure \ref{fig:ComparisonCouplings} shows a comparison of the couplings from the free free fermion case to the case $a\sim\eta_B>0$. Obviously the decay is much faster than $1/d$ thus resolving the divergence. This is best seen in the Fourier transform, obtained by the same calculation as above:

\begin{widetext}
   \begin{multline}
  \widetilde g_a(k)  =  -\frac{V^2}{16 \pi J_F}\sum_{l,C_1,C_2}\int_{0}^{\pi}{\rm d}\xi\,\frac{\Theta(\pi-2\pi l C_2+C_1 C_2\xi+C_2 k )\Theta(2\pi l C_2-C_1 C_2\xi-C_2 k )}{\sqrt{\cos^2(\xi)+a^2}+\sqrt{\cos^2(C_1 \xi+k)+a^2} }\\
  \left(1+ \frac{\cos(\xi)}{\sqrt{\cos^2( \xi)+a^2}}\right)\left(1-\frac{\cos( C_1 \xi+k)}{\sqrt{\cos^2( C_1 \xi+k)+a^2}}\right).\label{eq:ResultFourierTrafoRenormalized}
  \end{multline}
\end{widetext}

\begin{figure}[t]
  \centering
\Bild{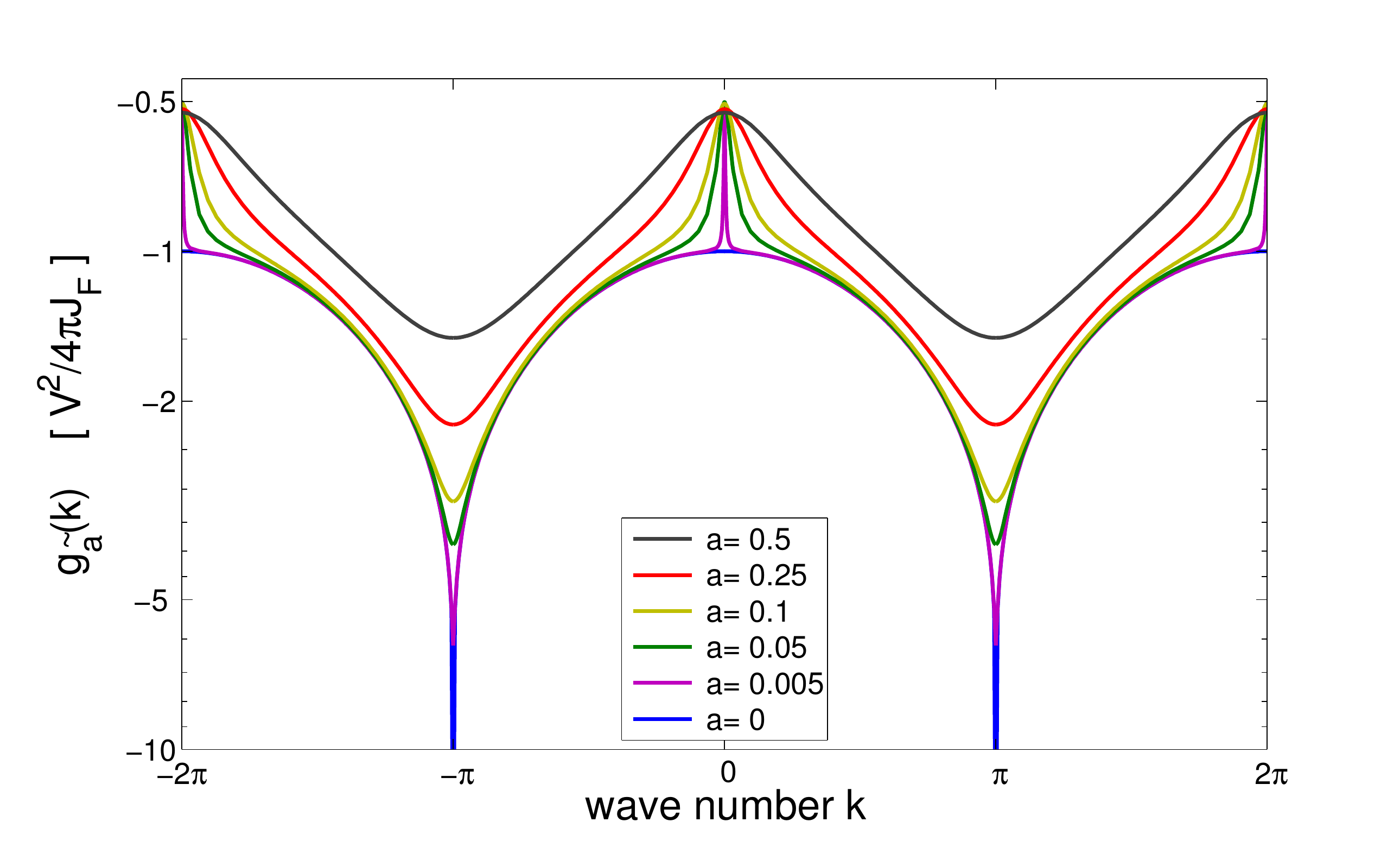}
     \caption{(Color online) Fourier transform of the renormalized couplings $\widetilde g_a(k)$ for different amplitude factors $a$. At $k=\pm\pi$, the divergence for $a\to0$ is clearly observable. For $k=0$ the situation is more complicated and a short discussion can be found in the main text.}
   \label{fig:FourierTransformedCouplingsRenormalized}
    \end{figure}

The numerical integration shown in figure \ref{fig:FourierTransformedCouplingsRenormalized} for different amplitude factors $a$ directly reveals the lifting of the divergence at $k=\pm\pi$. It should be noted that the limits $k\to0$ and $a\to0$ are not interchangeable and need to be treated carefully as can be seen from the figure. Relevant in our case is only the limit $a\to0$ for $k=0$ as discussed later on. For these most important cases $k=0,\pm\pi$, analytic expressions for the Fourier transformed couplings could be found. Here, a detailed analysis of the summations over $(l,C_1,C_2)$ reveals a very limited set of contributions, resulting in
  \begin{align}
    \widetilde g_a(\pm\pi)  &= -\frac{V^2}{8\pi J_F}\frac{1}{\sqrt{1+a^2}} \left(2K[\frac{1}{1+a^2}]-E[\frac{1}{1+a^2}]\right)\\
    \widetilde g_a(0)  &=-\frac{V^2}{8\pi J_F }\frac{1}{\sqrt{1+a^2}} E[\frac{1}{1+a^2}],
  \end{align}
with $E[x]$ being the complete elliptic integral of second kind \cite{Abramowitz1964}.\\

\subsection{The renormalized Hamiltonian}
The knowledge of the renormalized couplings finally allows to write down the effective bosonic Hamiltonian for half fermionic filling $\rho_F=1/2$. Starting from \eqref{eq:effectiveBHMFull} together with the renormalized fermionic density \eqref{eq:FermionicDensity}, the couplings \eqref{eq:CouplingsRenormalized} and the ansatz for the bosonic CDW \eqref{eq:AnsatzBosonicCDW}, the full effective bosonic Hamiltonian is given by
 \begin{multline}
  \H B^{\rm eff} =
-J_B\sum_j\left(\ad_j\a_{j+1}+\ad_{j+1}\a_{j}\right)+\frac{U}{2}\sum_j\nb_j\left(\nb_j-1\right)\\
  -\bar\mu\sum_j \nb_j-\Delta\sum_j\nb_j (-1)^j +\sum_j\sum_d g_d(a)\  \nb_j\nb_{j+d}\label{eq:effectiveHamiltonian}.
 \end{multline}
 Beside the usual hopping and interaction terms, two prominent features arise. On the one hand, the already discussed long-range density-density interaction with couplings  $g_d(a)$ lead to the emergence of CDW phases. These are further stabilized by the occurring induced alternating potential with amplitude
  \begin{equation}
  \Delta=2\rho_B\eta_B \widetilde g_a(\pi)+V \eta_F^a/2,\label{eq:InducedAlternatingPotential}
 \end{equation}
 being a direct consequence of the fermionic density wave
\begin{equation}
 \langle\nf_j\rangle_{\rm F}=\frac12\left[1-\eta_F^a (-1)^j\right].
\end{equation}
Though derived only for the case of double-half filling, the emergence of the induced chemical 
 \begin{equation}
  \bar\mu=2\rho_B \widetilde g_a (0)-V/2 \label{eq:InducedChemicalPotential}
 \end{equation}
 in combination with the general ansatz also allows for an extension of the effective Hamiltonian to other fillings $\rho_B$ as done later on. For completeness, the  amplitude factor $a=\frac{V\eta_B\rho_B}{2\hbar J_F}$ couples to the amplitude $\eta_B$ of the induced bosonic CDW which is still a free parameter. For the Fourier transform, the identities
\begin{equation}
 \sum_d g_d(a) = \widetilde g_a(0),\hspace{1cm} \sum_d (-1)^d\ g_d(a) = \widetilde g_a(\pi),
\end{equation}
hold. Before we exploit the resulting Hamiltonian in the determination of the phase diagram, possible approaches in a self-consistent determination of the bosonic CDW amplitude are shown in the next section.

\section{Self-consistent determination of \texorpdfstring{$\eta_B$}{etaB}}\label{chap:SelfconsistentAmplitude}
The introduction of the bosonic CDW amplitude $\eta_B$ as a free parameter demands for a proper way in the determination thereof. Although the knowledge of $\eta_B$ as a function of the bosonic hopping $J_B$ is not necessary in the discussion of the phase diagram as done in our approach, a possible reproduction of figure \ref{fig:CDWAmplitude} further supports the validity of our chosen approach. This prediction of the CDW amplitude on ground of Hamiltonian \eqref{eq:effectiveHamiltonian} is obtained from possible choices of the ground state together with a minimization of the resulting energy. Within the minimization scheme, the free parameter is obtained from analytic results for the energy.\\

\begin{figure}
  \centering
\Bild{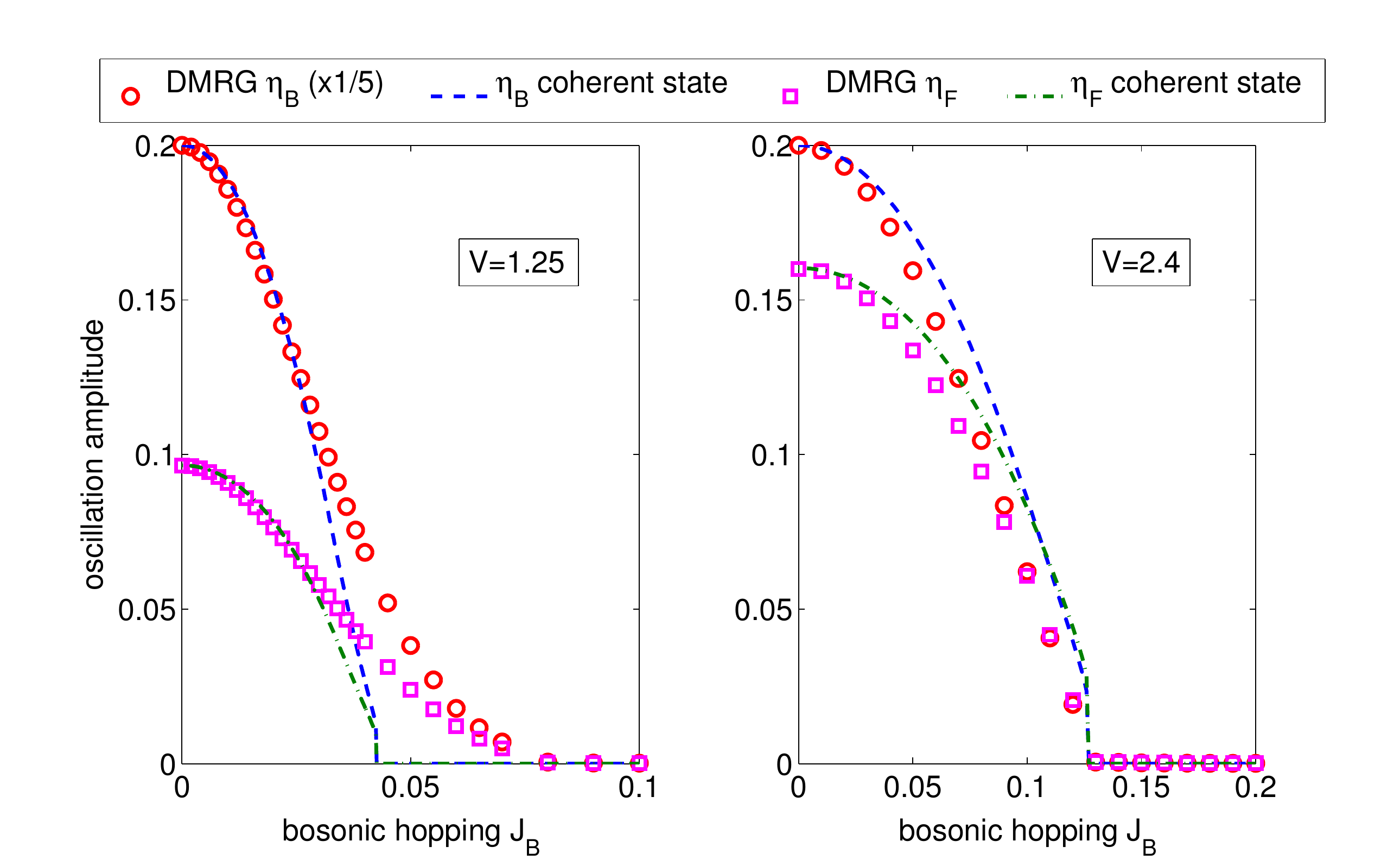}
 \caption{(Color online) Self-consistent determination of the amplitude of the bosonic CDW from the minimization of the energy for the effective Hamiltonian with respect to a coherent state ansatz. Shown are the same numerical results as in figure \ref{fig:CDWAmplitude} (left plot, $L=512$) as well as results for $V=2.4$ and $J_F=10$ (right, $L=256$). One can see the rather poor quantitative agreement with a general qualitative agreement.}
 \label{fig:SelfconsistentCoherentAmplitude}
\end{figure}

\textbf{Coherent state:}
The simplest choice for the ground state of Hamiltonian \eqref{eq:effectiveHamiltonian} is given by local coherent states $\left|\alpha\right\rangle$ \cite{Glauber1963}. Using the ansatz
\begin{equation}
\left| \Psi \right\rangle^{\rm coh} = \prod_{j=-\infty}^\infty   \left|\alpha_+\right\rangle_{2j}\left|\alpha_-\right\rangle_{2j+1},
\end{equation}
the requirement of proper local densities
\begin{equation}
 ^{\rm coh}\left\langle \Psi\right| \nb_j \left|\Psi\right\rangle^{\rm coh} = \frac12\left[1+\eta_B (-1)^j\right]
\end{equation}
as assumed in the derivation of the Hamiltonian lead to a direct connection between the coherent amplitudes $\alpha_\pm$ and $\eta_B$ as $\alpha_\pm=\sqrt{\frac12\pm \frac12\eta_B}$. The energy $E\left[\eta_B\right]=\ ^{\rm coh} \left\langle\Psi\right| \H B^{\rm eff} \left|\Psi\right\rangle^{\rm coh}$ now becomes a function of $\eta_B$ and upon neglecting unphysical contributions from the interaction \footnote{Although the coherent state incorporates all Fock states $n$, for the treated CDW only states with $n\le 1$ are of importance.}, the energy is given by
\begin{align}
E\left[\eta_B\right] &= -J_B \sqrt{1-\eta_B^2} +\frac12 g_0(a)\\
&\hspace{1cm}- \frac14V\eta_F^a \eta_B-\frac14 \widetilde g_a(\pi) \eta_B^2-\frac14 \widetilde g_a(0).\notag
\end{align}
We stress that the amplitude factor $a=\frac{V\eta_B}{4\hbar J_F}$ as well as the fermionic amplitude $\eta_F^a$ also depend on $\eta_B$. Minimization of this function with respect to $\eta_B$ at the end gives a prediction of the bosonic CDW amplitude. This is shown in figure \ref{fig:SelfconsistentCoherentAmplitude}, where the self-consistent prediction is compared to the numerical data from figure \ref{fig:CDWAmplitude} and to data for $V=2.4$ and $J_F=10$. One can see that the coherent approach gives a qualitatively good agreement for small $J_B$, but the quantitative agreement is rather poor in particular for larger $J_B$ because of the strongly simplified ansatz used here.\\

\begin{figure}
  \centering
\Bild{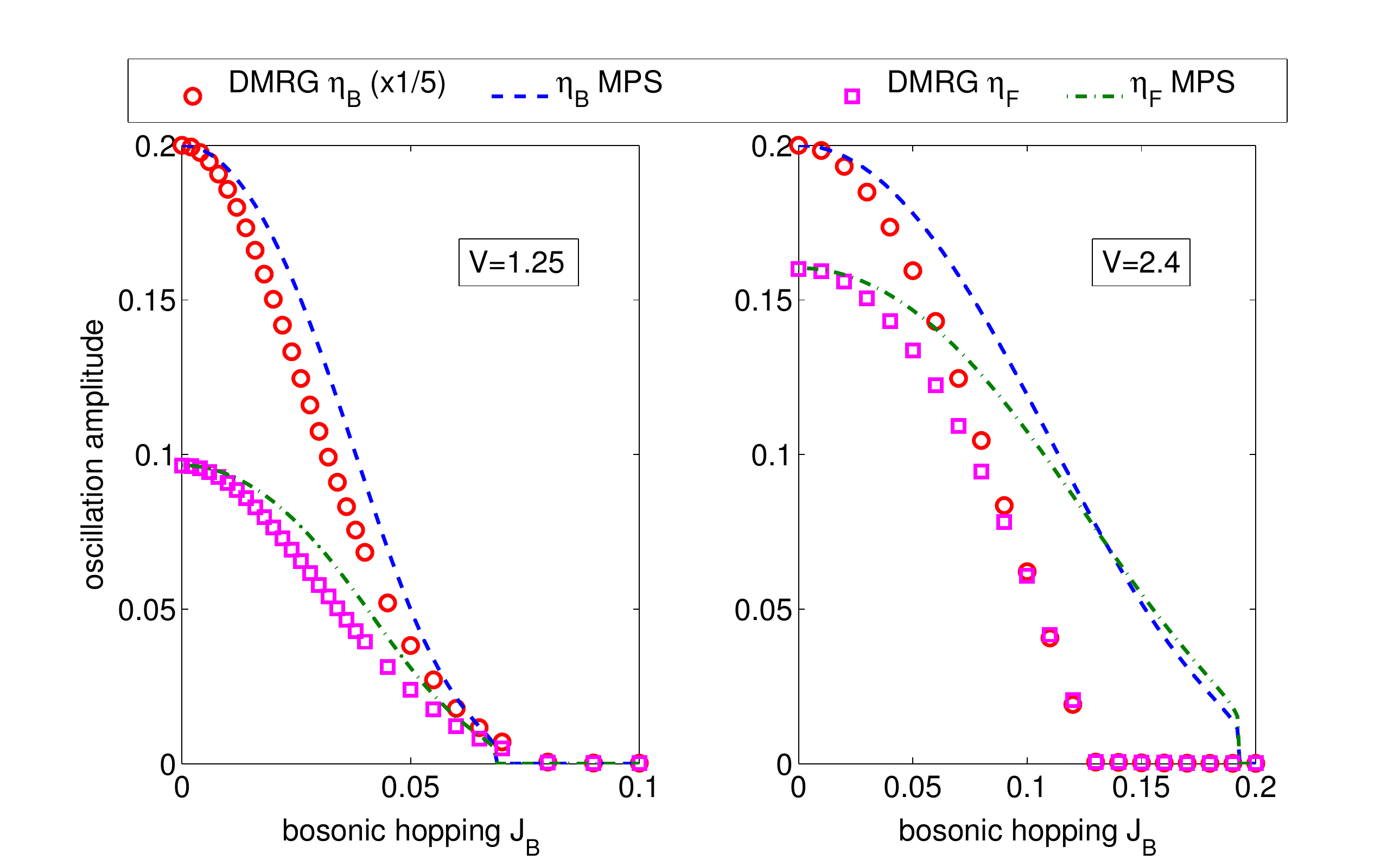}
 \caption{(Color online) Self-consistent determination of the amplitude of the bosonic CDW from the minimization of the energy for the effective Hamiltonian with respect to a matrix product state ansatz. Shown are the same numerical results as for figure \ref{fig:CDWAmplitude} (left plot, $L=512$) and results for $V=2.4$ and $J_F=10$ (right, $L=256$). One can see the better quantitative agreement compared to the result for the coherent state in figure \ref{fig:SelfconsistentCoherentAmplitude} for small amplitude factor $a$, i.e., for small interaction $V$.}
 \label{fig:SelfconsistentMPSAmplitude}
\end{figure}

\textbf{Matrix product state:}
Better results for the CDW amplitude may be found from a matrix product like ansatz. Using a different description of the two-site blocks by the ansatz
\begin{equation}
\left| \Psi \right\rangle^{\rm MPS} = \prod_{j=-\infty}^\infty   \sum_{i_1,i_2=0}^1 A_{i_1i_2}\ \left|i_1\right\rangle_{2j}\left|i_2\right\rangle_{2j+1},
\end{equation}
the problems arising from the higher number states are ruled out by definition. With the introduction of the prefactors $A_{i_1i_2}$ which are chosen to be real, we introduce four free parameters which have to be minimized in general.  This set of parameters can be reduced by constraints from the normalization of the ground state as well as the expected local densities \eqref{eq:AnsatzBosonicCDW}. Altogether, these constraint reduce to $A_{00}=A_{11}\equiv0$ and the energy functional only depends on $\eta_B$ as
   \begin{equation}
    \begin{split}
      E[\eta_B] &=  - J_B \sqrt{1-\eta_B^2} -\frac V2\eta_F^a \eta_B-\frac12 \widetilde g_a(\pi) \eta_B^2-\frac12 \widetilde g_a(0)\\
  &\hspace{1cm}+ (1-\eta_B^2)\bigl[\frac12 g_0(a)-\frac12 g_1(a)\bigr].
    \end{split}
   \end{equation}
 The corresponding numerical results for the minimization can be found in figure \ref{fig:SelfconsistentMPSAmplitude}. The quantitative agreement is slightly better compared to the coherent state approach for smaller interaction $V$ but still the strong simplification of the ansatz pays its tribute. For larger $V$, the made matrix product ansatz seems to fail. We believe this to be connected to the increasing induced alternating potential which fosters higher number states. Nevertheless, the two presented self-consistent determinations of the amplitude $\eta_B$ show that this free parameter in principle may be calculated with more sophisticated ansatzes involving higher number states. As will be seen in the next chapter, an exact calculation of $\eta_B$ as a function of the bosonic hopping is not of importance however.
 
\section{Phase diagram of the effective bosonic model}\label{chap:ResultsFastFermionsFullModel}
We now use the effective bosonic Hamiltonian 
 \begin{multline}
  \H B^{\rm eff} =
-J_B\sum_j\left(\ad_j\a_{j+1}+\ad_{j+1}\a_{j}\right)+\frac{U}{2}\sum_j\nb_j\left(\nb_j-1\right)\\
  -\bar\mu\sum_j \nb_j-\Delta\sum_j\nb_j (-1)^j +\sum_j\sum_d g_d(a)\  \nb_j\nb_{j+d}\label{eq:effectiveHamiltonian2}
 \end{multline}
to calculate the full phase diagram and compare it to the numerical results from figure \ref{fig:PhaseDiagramFastFermions}. As a reminder, the potentials $\bar\mu$ and $\Delta$ are given by 
 \begin{equation}
  \bar\mu=2\rho_B\widetilde g_a (0)-V/2, \hspace{1cm}  \Delta=2\rho_B\eta_B \widetilde g_a(\pi)+V \eta_F^a/2.
 \end{equation}
 The calculation of the phase boundaries of the different incompressible regions (MI, CDW) is performed from a canonical point of view, i.e., the particle-hole gap is analyzed. For an incompressible phase with filling $\rho_B$, the chemical potentials of the upper and lower boundaries are obtained from
 \begin{equation}
  \mu_{\rho_B}^\pm = \pm \Bigl[E(\rho_BL\pm1)-E(\rho_BL) \Bigr].
 \end{equation}
First we restrict ourselves to the zero-hopping limit $J_B=0$, whereas later on we employ a full degenerate perturbation theory in $J_B$. Concerning the bosonic amplitude $\eta_B$ it should be mentioned, that both, in the zero hopping limit (here the amplitude naturally equals one) as well as in the small hopping region, $\eta_B=1$. In the latter situation this is the case because the perturbation theory starts at $J_B=0$ and all energies and quantities are to be calculated for this case.

\subsection{Zero-hopping phase diagram}
The calculation of the chemical potentials for $J_B=0$ is straightforward. In this case, the energy is given by a replacement of the number operators $\hat n_j$ in \eqref{eq:effectiveHamiltonian2} by real numbers according to the ground state in the system. Additionally, the density $\rho_B$ and a possible CDW amplitude $\eta_B$ needs to be fixed, too. This is done for the Mott insulator with unity filling $(\rho_B;\eta_B;\left\langle\hat n_j\right\rangle)=(1;0;1)$, the empty Mott insulator $(0;0;0$) as well as the CDW $(\frac12;1;\frac12[1+(-1)^j])$ and the corresponding particle or hole states. Possible competing ground state configurations as $(1;1;1+(-1)^j)$ for the case of a Mott insulator are ruled out from a detailed analysis of the resulting energies. Also the different choices for the position of the additional particle (or hole) are considered. Finally, the configuration with minimal energy is used for the determination of the boundaries.

For the prediction of the phase diagram in the considered regime of chemical potentials $\mu_B$, this results in
 \begin{align}
 \mu^-_1 &=\frac V2 - g_0(0),\\
  \mu^\pm_{\frac12} &= \frac V2\pm\frac V2\eta_F^a \pm g_0(a),\\
   \mu^+_0 &= \frac V2 +  g_0(0)  , 
 \end{align}
  \begin{figure}[t]  
  \centering
 \Bild{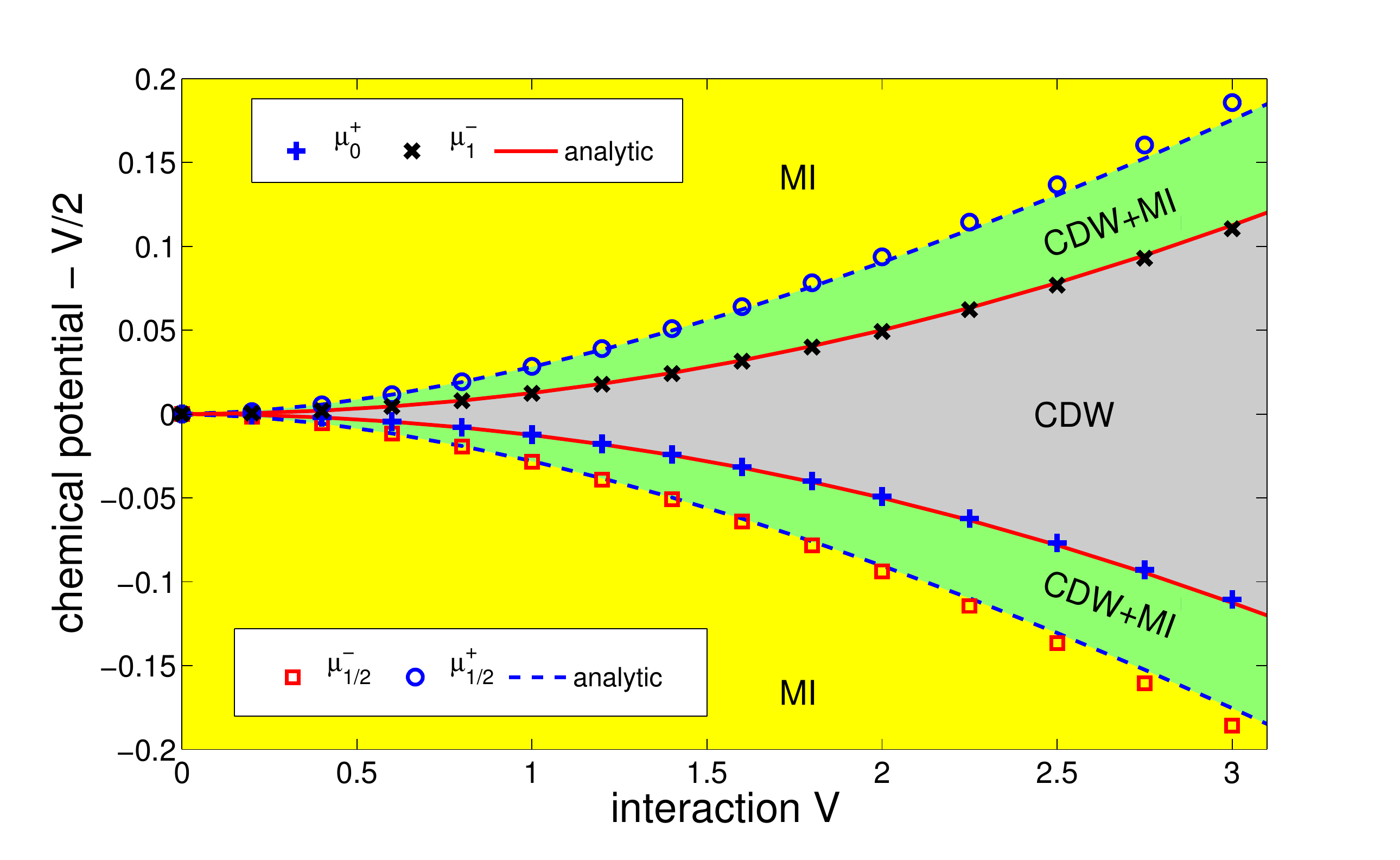}
   \caption{(Color online) Phase diagram of the effective bosonic Hamiltonian for vanishing bosonic hopping $J_B=0$. Data points are the numerical results obtained from DMRG and ED for the full BFHM and the lines are the analytic results. Yellow: extended region of the different Mott insulators (MI). Gray: charge density wave phase (CDW); Green: Coexistence region of CDW and MI. Fermionic density $\rho_F=1/2$ and $J_F=0$.}
    \label{fig:ZeroHoppingPhaseDiagram}
    \end{figure}
 which together with the results for the couplings $g_d(a)$ \eqref{eq:CouplingsRenormalized} and the fermionic CDW amplitude $\eta_F^a$  from \eqref{eq:FermionicDensity} allow to construct the phase diagram at zero bosonic hopping. This is shown in figure \ref{fig:ZeroHoppingPhaseDiagram}, where the chemical potentials are displayed as a function of the interaction $V$ for a fixed fermionic hopping $J_F$, cleaned up from the mean-field shift $\frac V2$. 
 
 The first observation from the figure is the very good agreement of the numerical results for the full BFHM and the analytic results obtained from the effective bosonic Hamiltonian. Increasing deviations for larger $V$ could be addressed both to the breakdown of the Markov approximation \eqref{eq:MarkovApproximation} as well as the negligence of higher order contributions in \eqref{eq:CumulantRenormalizedFermions}. Most prominent feature is the overlap between the MI and CDW phases, i.e., $\mu_0^+> \mu_{\frac12}^-$ and $\mu_1^-< \mu_{\frac12}^+$. This behavior, already seen in figure \ref{fig:PhaseDiagramFastFermions},  is uncommon since it indicates a \emph{negative compressibility}
 \begin{equation}
 \kappa = \frac{\partial \langle\hat N\rangle}{\partial \mu_B} < 0
 \end{equation}
 within the coexistence phase. These kind of coexistence phases are not new (see e.g. \cite{Batrouni2000,Titvinidze2008,Hubener2009,Soeyler2009} for a variety of different coexistence phases), but the coexistence of a Mott insulator and a CDW phase has to our knowledge not been reported before. Though being uncommon, a physical explanation of this effect could easily be given. In the grand-canonical ensemble, the coexistence phase does not exist since in this situation, the number of particles in the system is chosen such that the energy is minimized: this drives the system always into a CDW phase within this region. From a canonical point of view, adding further particles to the CDW phase results in configurations, where the repulsive contribution to the energy remains constant whereas the attractive one is increased; the energy per particle is thus reduced. This of course holds in the thermodynamic limit. For finite systems, the boundaries play a vital role as discussed in section \ref{sec:Boundaries}. Within the coexistence phase in a canonical picture, the resulting ground state displays a configuration, where one part of the lattice inherits all features of a MI and the other part those of a CDW. This corresponds to a phase separation between MI and CDW. 
 
Finally, some remarks on the data points in figure \ref{fig:ZeroHoppingPhaseDiagram}. These are obtained from numerical results, where the Mott insulators are calculated using a finite size extrapolated exact diagonalization and the numerical results for the CDW are resulting from DMRG calculations where the boson distribution is fixed, acting as a potential to the fermions. This procedure is necessary here, since the full DMRG for this system has severe problems in obtaining the proper ground state. The reason for this is on the one hand the sensitivity of the system to the boundary in the open boundary DMRG and on the other hand the problem of seeking the ground state within the energy manifold with many close-lying meta-stable states. This complicates the numerical calculation enormously. A detailed discussion of the boundary issue can be found in  section\ref{sec:Boundaries}.

\subsection{2nd order strong-coupling theory}
Going beyond the zero-hopping limit, we aim at a perturbative treatment in the hopping amplitude $J_B$. This allows to generate the full phase diagram in the $(\mu_B,J_B)$ plane. Since the methodology of the perturbation theory is quite involved, we only present the basic ideas. Details could be found in \cite{ThesisMering2010}.

Basic idea of the degenerate perturbation theory is the determination of the ground-state energy in second order in the bosonic hopping. Since the particle (hole) states obtained from adding (removing) a particle to (from) the system are highly degenerate, degenerate perturbation theory has to be applied within the degenerate manifold of states $\ket\Psi_l$. The index just labels the different states within the manifold. Different formulations of degenerate perturbation theory exist (e.g., as \cite{Freericks1996}  used in  for the pure and disordered Bose Hubbard model), where we use Kato's expansion \cite{Klein1974,Teichmann2009,Eckardt2009}, which relies on the calculation of an effective Hamiltonian (in arbitrary order) within the degenerate subspace. The last step is to solve this effective Hamiltonian and obtain the ground-state energy as a function of the perturbation parameter. 

Up to second order, Kato's expansion is given by
\begin{equation}
 \hat H^{\rm eff} = E_0+\mathcal P \hat H_1 \mathcal P + \mathcal P \hat H_1\mathcal Q\frac{1}{E_0-\hat H_0}\mathcal Q \hat H_1\mathcal P,\label{eq:Kato}
\end{equation}
where $\mathcal P$ is the projector onto the degenerate subspace, $\mathcal Q=\mathbbm 1-\mathcal P$ the orthogonal projector and $E_0$ is the zero order energy of the manifold. Here, the Hamiltonian is written in the form $ \hat H = \hat H_0 + \hat H_1$, where $\hat H_1$ is the perturbation, i.e., the hopping in our case. For the calculation of the effective Hamiltonian, only the action of \eqref{eq:Kato} on any input state $\ket \Psi_l$ from the degenerate subspace needs to be studied. In our case, the result is of the form
\begin{align}
 \hat H^{\rm eff} \ket\Psi_l &= E_0(\ket\Psi_l) + J_1\Bigl[ \ket\Psi_{l-1}+\ket\Psi_{l+1}\Bigr]\\
 &\hspace{1cm}+ J_2 \Bigl[ \ket\Psi_{l-2}+\ket\Psi_{l+2}\Bigr]+W \ket\Psi_l\notag
\end{align}
since our perturbation only consists of a nearest-neighbor hopping. In \cite{ThesisMering2010}, a generalization to arbitrary long-range hopping can be found. This (maximally) tridiagonal matrix representation of the effective Hamiltonian can be solved by a Fourier transform, which gives the energy
\begin{equation}
 E = E_0 + 2 J_1 \cos(2\pi\frac{k}{L}) + 2 J_2 \cos(4\pi\frac{k}{L}) +W,
\end{equation}
where the $k$ mode has to be chosen such that the energy is minimal. In this system this is typically the case for $k=0$ since both $J_1\sim J_B$ and $J_2\sim\frac{ J_B^2}{E_0-\langle\hat H_0\rangle}$ are negative. A crucial point in the calculation comes from the nature of the effective bosonic Hamiltonian in \eqref{eq:effectiveHamiltonian2}. Since the density-density interaction is long ranged, the energy denominator depends on the distance of the particle performing the first hopping process from the reference site where the additional particle (hole) is situated. This needs to be taken into account for the calculation of the chemical potentials. Since the expressions for the chemical potentials are lengthy, these are only given in appendix \ref{sec:AppendixChemicalPotentials}.

\begin{figure}[t]  
  \centering
 \Bild{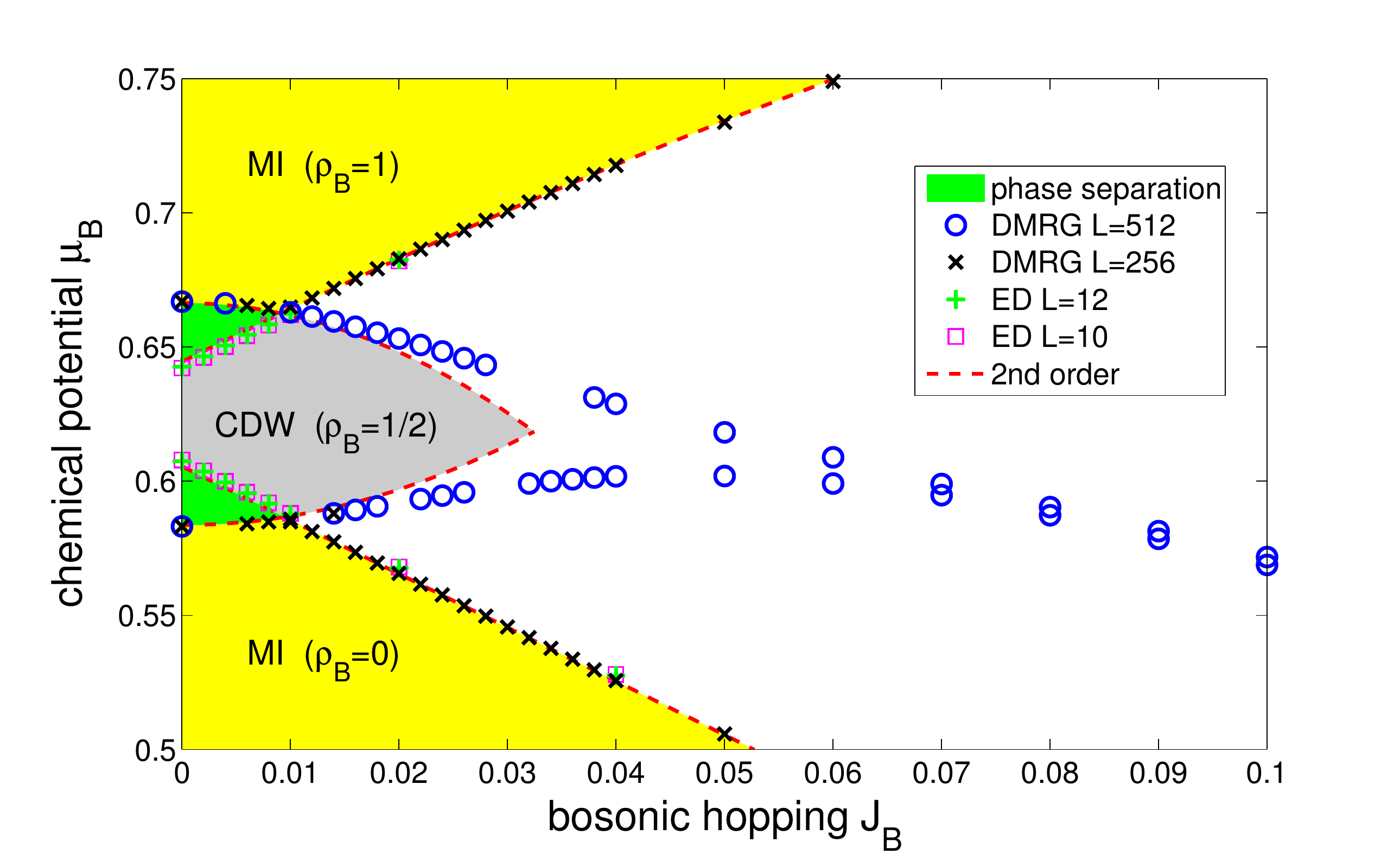}
   \caption{(Color online) Analytic results for the phase diagram together with the numerical results from figure  \ref{fig:PhaseDiagramFastFermions}. The agreement between the analytics and the numerics is quite reasonable with the natural deterioration for larger hopping $J_B$ due to the perturbative treatment..}
    \label{fig:PhaseDiagramFastFermionsAnaNum}
    \end{figure}

A major difficulty in the calculation of the chemical potentials is the dependence of the results on all coupling strengths $g_d(a)$, which need to be calculated up to a large distance. For the analytic results used in figure  \ref{fig:PhaseDiagramFastFermionsAnaNum} it turns out, that $d\approx100$ is sufficient to gain convergence. Here we only give the numerical values for the chemical potential. Using \eqref{eq:ChemicalPotentialZero}-\eqref{eq:ChemicalPotentialUnity} and directly plugging in numbers, these are given by
\begin{align}
 \mu_0^+ &=0.605469-2 J_B,\\
 \mu_{\frac12}^-&=0.583612+33.076 J_B^2,\\
 \mu_{\frac12}^+&=0.666388- 45.4392 J_B^2,\\
 \mu_1^-&=0.644531+2 J_B-4.12927 J_B^2
\end{align}
Figure \ref{fig:PhaseDiagramFastFermionsAnaNum} shows the previously used numerical data from figure \ref{fig:PhaseDiagramFastFermions} together with the analytic predictions. From the figure it can be seen, that the overall agreement is quite reasonable compared to a second order treatment. Altogether, our analytic approach allows to completely derive the bosonic phase diagram analytically up to a good agreement and provides an intuitive physical understanding of the arising effects.

\subsection{Boundary effects in an effective model with long-range interactions}\label{sec:Boundaries}
As already mentioned at several places, boundary effects play an important role in this system. The long-range character of the fermion mediated interactions leads to a substantial modification of the system dynamics even for relatively large systems. This can directly be seen for the case of the CDW phase, where we first discuss the zero hopping case. Adding a further particle to the CDW phase, this particle has to choose an odd side. Since due to the open boundaries the translational symmetry is broken, it matters whether the particle is added close to the boundary or at the center. Because of the long-range interaction, the possible choices differ in energy, where the additional energy close to the boundary is given by $\sum_{d=-L/4}^{L/4} g_{2d+1}$, in contrast to the energy at the center $ \sum_{d=0}^{L/2} g_{2d+1}$. According to the properties of the couplings, the energy is minimal for a position close to the boundary. Adding further particles, the same arguments apply and whilst increasing the filling, a Mott insulating region is growing from the boundary. Switching to small, but finite hopping does not change the situations. As long as the hopping is small compared to the energy difference between the state with a particle pinned close to the border and the state with the additional particle at the center, the reduction of the interaction energy due to the pinning to the boundary dominates the increase of the kinetic energy. When removing a particle from the system, i.e., going below half filling, the same arguments apply.

  \begin{figure}[t]  
  \centering
 \Bild{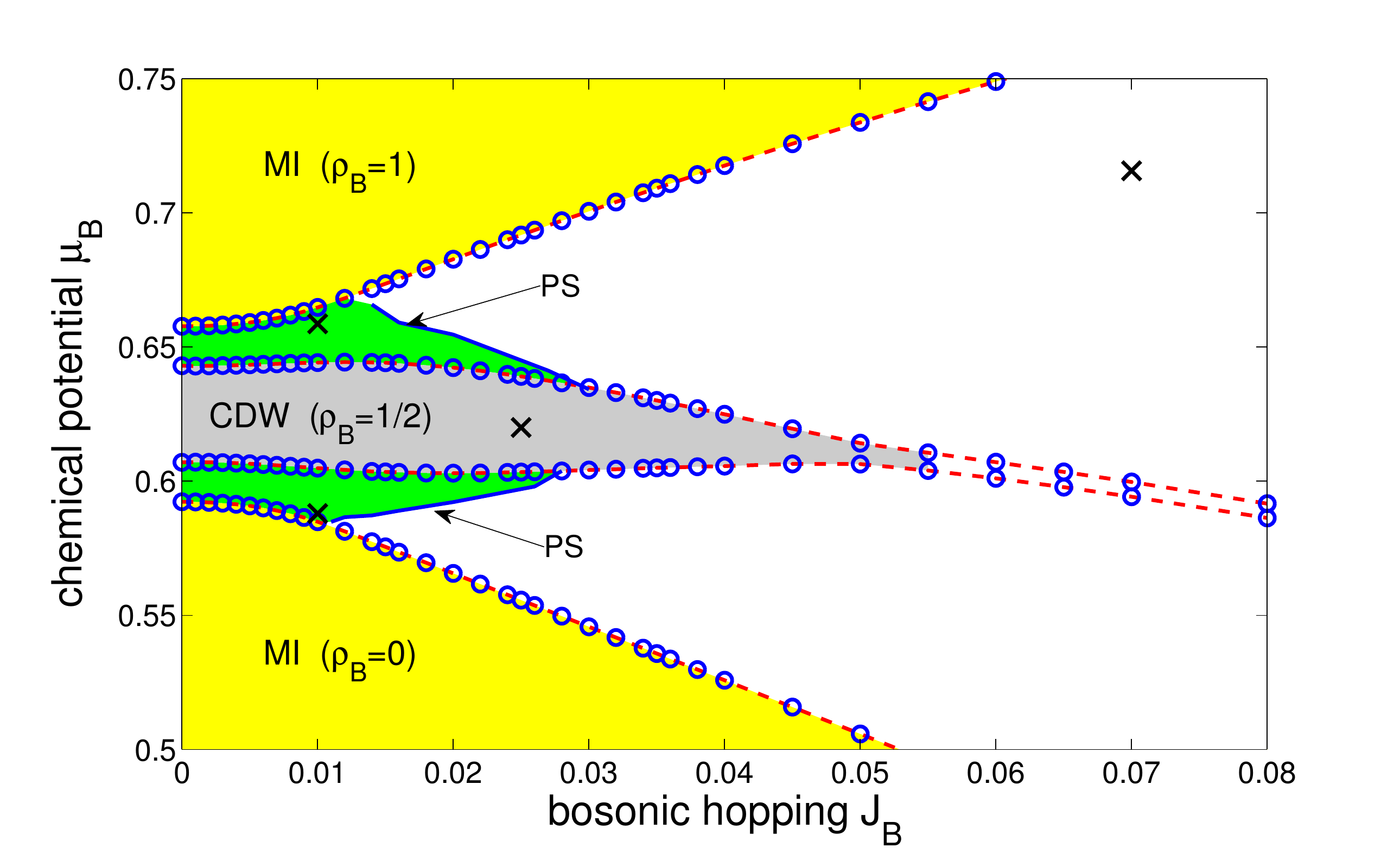}
   \caption{(Color online) Phase diagram of the full BFHM with open boundaries. One can see that the lobes bend apart from each other, resulting in an extent region where the CDW and the Mott insulator exist together (filled green region), but with a spatial phase separation (PS). The data are obtained with DMRG and open boundary conditions for a fixed length of $L=128$ sites. The other parameters are $J_F=10$ and $V=1.25$. Black crosses show the points where the density profiles in figure \ref{fig:DensityCutsFastFermions} are taken from. The dashed lines are to guide the eye.}
    \label{fig:NumericalResultsBoundary}
    \end{figure}
  
\begin{figure}[t]  
  \centering
  \Bild{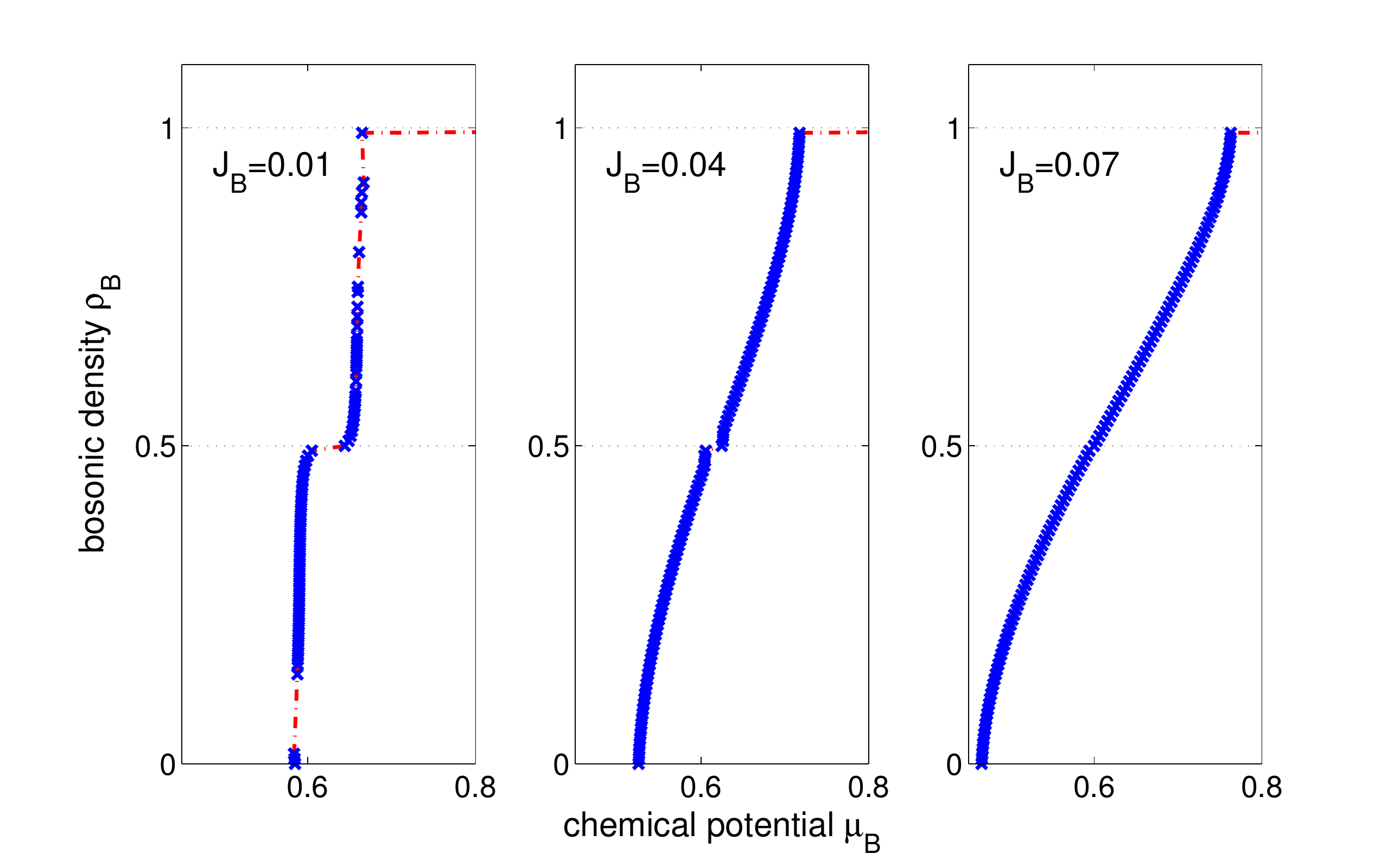}
    \caption{(Color online) Density cut along the chemical potential axis in figure \ref{fig:NumericalResultsBoundary}  for three different bosonic hoppings $J_B=0.01,0.04,0.07$ (from left to right). Shown is the density as a function of the chemical potential. Clearly the CDW and the Mott plateaus are visible, indicated by the extend region of constant filling, where there is no {\it devil's staircase} as might be expected because of the long-range interactions. For $J_B=0.01$, some data points are missing due to some convergence problems of the DMRG.}
    \label{fig:DensityCutLargeHopping}
    \end{figure}

  This behavior supports our observation of a \emph{phase separation between a Mott insulator and a CDW} in the infinite system with negative compressibility. However, the (open) boundary leads to a different dependence of the compressibility, now being strict positive$\kappa  > 0$. This can be seen from figure \ref{fig:NumericalResultsBoundary}, where the DMRG  results for a system exposed to open boundaries are shown. In contrast to \ref{fig:PhaseDiagramFastFermionsAnaNum}, the Mott lobes and the CDW phase bend apart from each other, not overlapping anymore. This is due to the positive compressibility due to boundaries. The positive compressibility could also be seen in figure \ref{fig:DensityCutLargeHopping}, where the bosonic filling is shown for three different cuts at fixed $J_B$ along the $\mu_B$-axis is shown. The filling is in each situation a monotonous function of $\mu_B$ according to $\kappa>0$. The incompressible CDW and MI phases are clearly observable. Interestingly our system dose not display a so-called {\it Devil's staircase} as described in \cite{Capogrosso-Sansone2010,Bak1982,Burnell2009} for the case of a dipolar Bose gas with density-density interactions decaying as $g_d\sim\frac{1}{d^3}$. Most likely, this is because of the alternating sign in our coupling constants together with the alternating potential, where a detailed discussion of this fact might be an interesting supplement to the present work.

\begin{figure}[t]  
  \centering
  \Bild{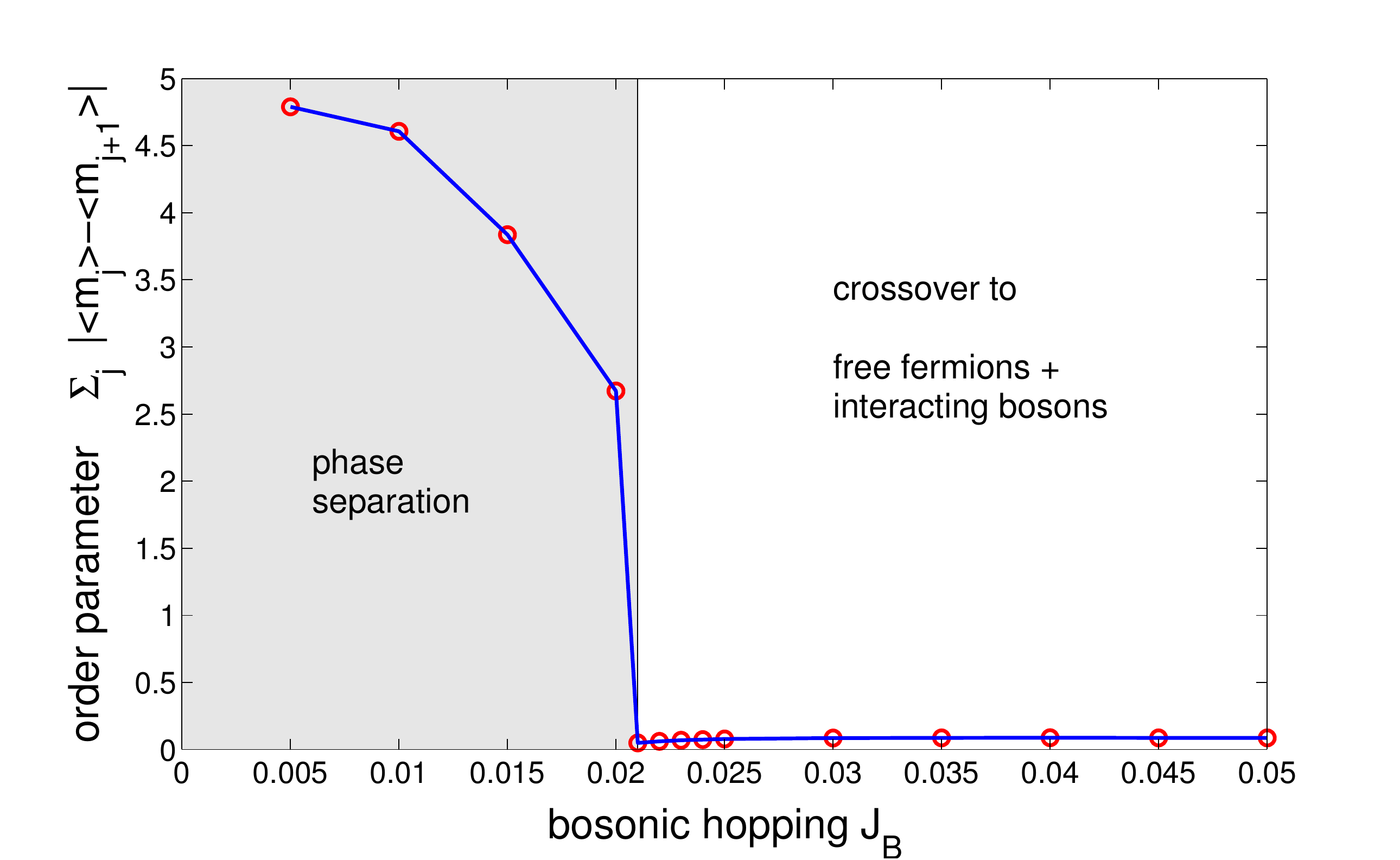}
   \caption{(Color online) Determination of the boundary of the phase separation using the order parameter $\mathcal O=\sum_j \left|\langle\hat m_j\rangle-\langle\hat m_{j+1}\rangle\right|$ for open boundary conditions as function of the bosonic hopping. The transition from the phase separation to the crossover regime is seen by the non-analyticity of the order parameter. Data points are obtained for $V=1.25$, $J_F=10$ and $24$ bosons on $64$ sites using DMRG for the full BFHM.}              
    \label{fig:OrderParameter}    
    \end{figure}

\begin{figure}[t]
  \centering
 \Bild{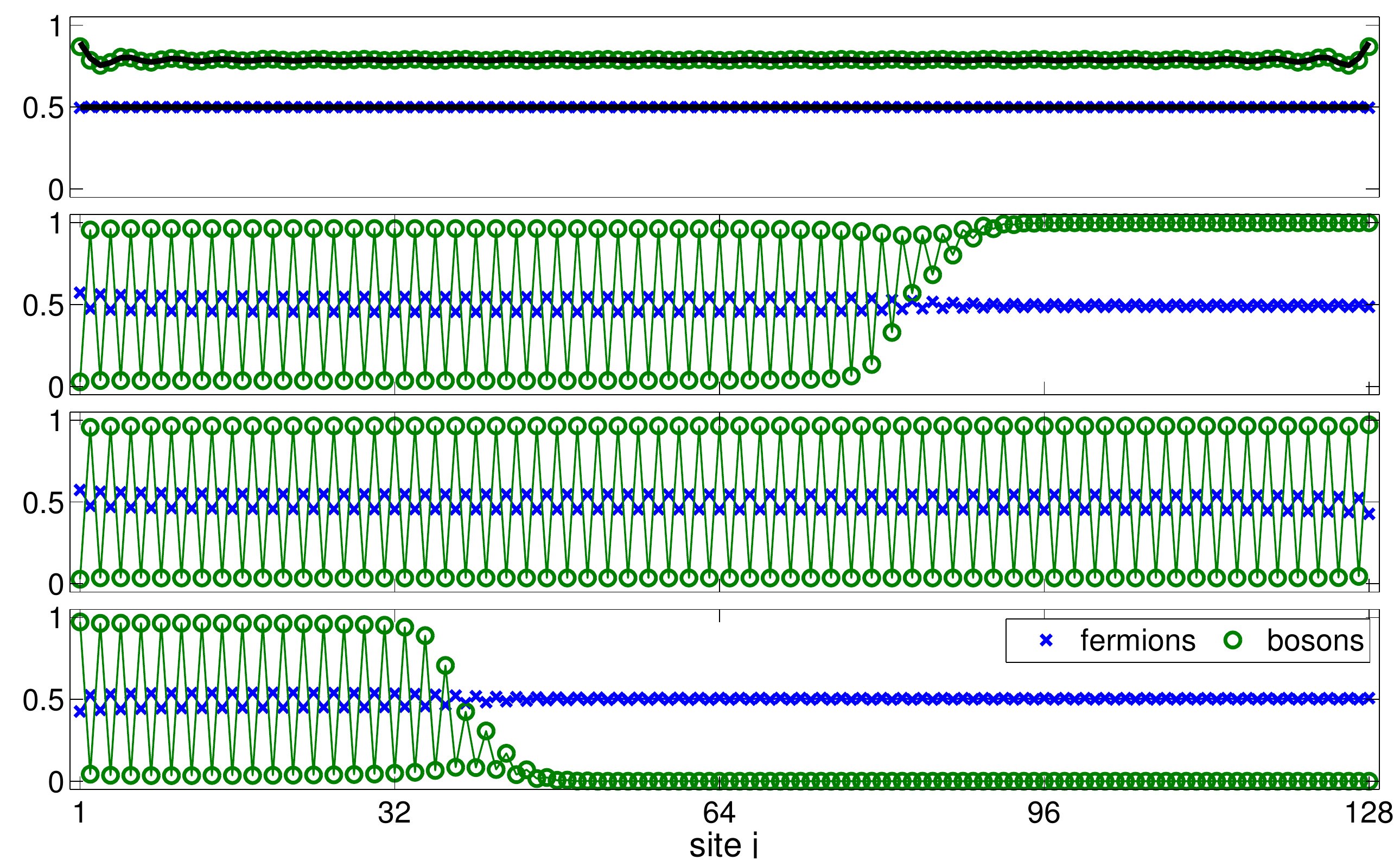}
    \caption{(Color online) Density profile obtained by DMRG for various numbers of particles. From bottom to top: $N_B=20,64,86,101$ for a system of $L=128$ sites. The lower three are for $J_B=0.01$ and the uppermost for $J_B=0.07$. One can immediately see the pinning of the additional particles to the boundary resulting in a phase separation of Mott insulator and CDW. In the uppermost plot, the fermionic state is roughly given by a homogeneous distribution according to the Friedel oscillations whereas the bosons behave as interacting bosons. This can be seen from the additional solid line which, gives the density profile for the same choice of parameters but without interspecies interaction, decoupling the bosons and the fermions. The positions of the data set for the density cuts in the phase diagram are depicted by the small marks in figure \ref{fig:NumericalResultsBoundary}.}
    \label{fig:DensityCutsFastFermions}
    \end{figure}

  For the phase diagram with open boundaries as presented in \ref{fig:NumericalResultsBoundary}, the extend of the phase-separated phase (PS) is sketched without a rigorous numerical analysis for rather small systems. The boundaries are determined from the behavior of the order parameter
\begin{equation}
\mathcal O=\sum_j \left|\langle\hat m_j\rangle-\langle\hat m_{j+1}\rangle\right|,
\end{equation}
which accounts for the CDW amplitude of the fermionic subsystem. Figure \ref{fig:OrderParameter} shows the behavior of the order parameter as a function of the bosonic hopping for $N_B=24$ and $L=64$. The phase boundary is clearly visible from the sharp drop around $J_B=0.02$. For larger hopping, the system enters a phase where the fermions finally behave as free fermions, i.e., showing a homogeneous density and the bosons have the density profiles according to that of interacting bosons (in the finite system). This behavior, together with examples from the PS and the CDW phase are shown in figure \ref{fig:DensityCutsFastFermions}. Here, for certain choices of the different parameters in the full BFHM, the resulting local density of bosons and fermions are shown. The phase separation between MI and CDW is evident.

\section{Conclusion and outlook}
Deriving an effective bosonic Hamiltonian we provided a comprehensive understanding of the bosonic phase diagram in the limit of ultrafast fermions. For double half filling, the physics is dominated by induced long-range density-density interactions alternating in sign, leading to the emergence of a bosonic charge-density wave phase. Divergences arising from the full decoupling of the fermions are overcome by a renormalization scheme which includes the back-action of the bosonic CDW on the fermions. Beyond half filling, the induced interactions lead to thermodynamically unstable regions in the $(\mu_B,J_B)$-phase diagram, displaying coexistence of CDW and Mott insulating phases, i.e., a phase separation between CDW and Mott insulator. Numerical results obtained by DMRG for the full BFHM are in a reasonable agreement with our analytic predictions. Application of our effective theory to Bose-Bose of Fermi-Fermi mixtures is straightforward.\\

Mainly focussing on the study of the incompressible and the phase-separated phases, the nature of the phase transition or crossover for double half filling remains open. Exponentially decaying first-order correlations even beyond the numerically detectable extent of the CDW phase indicate further physical processes in this system for larger bosonic hopping. Here bosonization could give an understanding of the behavior of the correlation functions as well as the nature of the phase transition. Furthermore, the question of supersolidity in the effective model is yet unanswered, where the interplay of the induced potential and the long-range interactions could lead to new effects. Focussing on low densities $\rho_B<1$, the nature of the different phases for larger boson densities is not studied so far and should give a variety of further phases.
\section{Acknowledgement}
This work has been supported by the DFG through the SFB-TR 49 and the GRK 792. 
We also acknowledge the computational support from the NIC at FZ J{\"u}lich and thank 
U. Schollw\"ock for his DMRG code. Furthermore we thank B. Capogrosso-Sansone, S. Das Sarma, E. Altmann, W. Hofstetter, C. Kollath, M. Snoek and T. Giamarchi for useful discussions.

\section*{Appendix}
\begin{appendix}
\section{Fourier transform of the coupling constants} \label{sec:AppendixFastFourier}
As discussed in section \ref{chap:FreeFermions}, the Fourier transform of the couplings (\ref{eq:CouplingsFreeFermions}) is governed by the Fourier transform of the numerator. At this place we will prove the result given in the main text using the Poisson sum formula \cite{Weisstein}
\begin{equation}
 \sum_{d=-\infty}^\infty f(d)= \sum_{l=-\infty}^\infty \int_{-\infty}^\infty f(x) e^{- 2\pi i l x}{\rm d}x.
\end{equation}
Rewriting of the cosine parts as
\begin{align}
 \sum_{d=-\infty}^\infty \cos d\xi \cos d\xi^\prime e^{i k d} &= \frac 14\sum_{C_1,C_2=-1,1}  \sum_{d=-\infty}^\infty  e^{i d\alpha}
\end{align}
with $\alpha=C_1 \xi + C_2 \xi^\prime + k$, application of Poisson's sum to the last term gives
\begin{align}
 \sum_{d=-\infty}^\infty e^{i \alpha d} &=   \sum_{l=-\infty}^\infty \int_{-\infty}^\infty e^{i x \alpha} e^{- 2\pi i l x}{\rm d}x\\
  &= 2 \pi  \sum_{l=-\infty}^\infty \delta(\alpha- 2\pi l).
\end{align}
Together with the definition of $\alpha$ we end up in the stated relation
\begin{multline}
 \sum_{d=-\infty}^\infty \cos d\xi \cos d\xi^\prime e^{i k d} =\\
 \frac\pi 2 \sum_{l=-\infty}^\infty \sum_{C_1,C_2=-1,1} \delta(2\pi l - C_1 \xi - C_2 \xi^\prime-k).
\end{multline}

\section{Fourier transform of the Green's functions}\label{sec:AppendixFourierTrafoGreens}
In section \ref{sec:SolutionDysonEquation}, the solution of the Dyson equation for the Green's function is presented. At this point we summarize the important points in the Fourier transform of the Green's functions, going back from the frequency domain to time domain. From equations (\ref{eq:GreensSolutionMomentumFrequencyEqual}) and (\ref{eq:GreensSolutionMomentumFrequencyShifted}) and with the definition of the Fourier transform
\begin{equation}
\mathcal G^{(\pm)}_{kk^\prime}(t+T,t)= \frac{1}{\sqrt{2\pi}}\int_{-\infty}^\infty {\rm d}\omega\, \mathcal G^{(\pm)}_{kk^\prime}(\omega)e^{i\omega T},
\end{equation}
the calculation of $\mathcal G^{(\pm)}_{kk^\prime}(t+T,t)$ is straight forward.\\

For $\mathcal G^{(\pm)}_{kk}(t+T,t)$, the Fourier transformation together with (\ref{eq:GreensSolutionMomentumFrequencyEqual}) gives
\begin{multline}
\mathcal G^{(\pm)}_{kk}(t+T,t)=\\
\pm\frac{i}{2\pi} \int_{-\infty}^\infty {\rm d}\omega\,  \frac{\epsilon_k\pm\omega\oplus i\delta}{\epsilon_k^2\oplus 2i\delta\epsilon_k-\delta^2-\omega^2+\frac{V^2\eta_B^2\rho_B^2}{\hbar^2}}\ e^{i\omega T}.
\end{multline}
Since the convergence factor $\delta$ is chosen in the limit $\delta\to0$, the $\delta^2$ in the denominator may be neglected. Looking at the dispersion of the free fermions and taking into account the definition of $\oplus$, the combination $\oplus \epsilon_k$ is always of negative sign, since for $k\in\mathcal K_F$, $\oplus$ means $+$ but $\epsilon_k<0$. In the other case the signs are just the other way around. Neglecting the factor of $2$ before the $\delta$ and defining the renormalized dispersion
 \begin{equation}
  \bar\epsilon_k=\sqrt{\epsilon_k^2+\frac{V^2\eta_B^2\eta_B^2}{\hbar^2}},
 \end{equation}
 the Green's function calculates as
 \begin{equation}
\mathcal G^{(\pm)}_{kk}(t+T,t)=\pm\frac{i}{2\pi} \int_{-\infty}^\infty {\rm d}\omega\,  \frac{\epsilon_k\pm\omega\oplus i\delta}{\bar\epsilon_k^2-\omega^2- i\delta}\ e^{i\omega T}.
\end{equation}
This integration is done by means of residue integration \cite{Jaenich1993}, where the contour is closed in the upper half plain, enclosing the pole at $\omega_0=-\sqrt{\bar \epsilon_k^2-i\delta}$. Finally, after performing the limit $\delta\to0$, the Green's function is given by
\begin{equation}
 \mathcal G^{(\pm)}_{kk}(t+T,t) = \frac{1 }{2}e^{-i \bar\epsilon_k T}\left(1\mp\frac{\epsilon_k}{\bar\epsilon_k}\right).
\end{equation}

In the unequal momentum case, the calculation for $\mathcal G^{(\pm)}_{kk+\frac L2}(t+T,t)$ follows the same route as described above. Using the result for the Green's function in the frequency domain (\ref{eq:GreensSolutionMomentumFrequencyShifted}) and the same argument for the combination $\oplus\epsilon_k$ as above, the Fourier transform is calculated from
\begin{equation}
 \mathcal G^{(\pm)}_{k\pm\frac{L}{2}k}(t+T,t)=i\frac{V\eta_B\rho_B}{2\pi\hbar} \int_{-\infty}^\infty {\rm d}\omega\, \frac{1}{\bar\epsilon_k^2-\omega^2- i\delta}e^{i\omega T}.
\end{equation}
Again closing the contour in the upper half plain with the same pole as above, the residue integration gives
\begin{equation}
 \mathcal G^{(\pm)}_{k\pm\frac{L}{2}k}(t+T,t) =-\frac{V\eta_B\rho_B}{2\hbar}\frac{1}{\bar\epsilon_k}e^{-i \bar\epsilon_k T}
\end{equation}
as stated in the main text.

\section{Chemical potentials within the perturbative treatment}\label{sec:AppendixChemicalPotentials}
As discussed in the main text, the derivation of the energies of the different particle-number states are determined in second order perturbation theory. The resulting chemical potentials read as:
\begin{widetext}
\begin{align}
 \mu_0^+ &= \frac V2 + g_0(0) -2 J_B\label{eq:ChemicalPotentialZero}\\
 \mu_{\frac12}^- &= \frac V2-\frac V2 \eta_F^a-g_0(a)+2 J_B^2\left(\frac{1}{V \eta_F^a+2g_{0}(a) -4g_{1}(a)+2 g_{2}(a)}-\frac{1}{V \eta_F^a+4g_{0}(a) -4g_{1}(a)}\right.\notag\\
   &\hspace{1cm}-\left.\sum_{m\  {\rm even}}\Biggl[\frac{1}{V \eta_F^a+2g_0(a)-2g_{1}(a)  }- \frac1{V \eta_F^a+2g_{0}(a) -2g_{1}(a)+2 g_{m}(a)-2 g_{m+1}(a)}\Biggr]\right)\\
  \mu_{\frac12}^+ &= \frac V2+\frac V2 \eta_F^a+g_0(a)-2J_B^2\left(\frac{4}{U+V \eta_F^a+4g_0(a)-4g_1(a)}\right.\notag\\
 &\hspace{0.5cm}- \frac{1}{V \eta_F^a+4 g_0(a)-4g_1(a)}+\frac{1}{V \eta_F^a+2 g_0(a)-4g_1(a)+ 2 g_{2}(a)}+ \frac{2}{U-V \eta_F^a}\\
 &\hspace{0.5cm} \left.-\sum_{m\ {\rm even}} \Biggl[\frac{1}{V \eta_F^a+2g_0(a)-2g_{1}(a)  }-\frac{1}{V \eta_F^a+2 g_0(a)-2g_1(a)+ 2 g_{m}(a)-2g_{m-1}(a)} \Biggr] \right)\notag \\
 \mu_1^- &= \frac V2 - g_0(0)+2 J_B-4 J_B^2\left( \frac{1}{U}+\frac{1}{U +4g_0(0)-4g_{1}(0)}-\frac{1}{U +2g_0(0)-4g_{1}(0)+2g_{2}(0)}\right.\label{eq:ChemicalPotentialUnity}\\
 &\hspace{1cm}\left.+\sum_{m}\Biggl[\frac{1}{U+2 g_0(0)-2g_1(0) }-\frac{1}{U +2g_0(0)-2g_{1}(0)+2g_{m+1}(0)-2 g_{m}(0)}\Biggr]\right)\notag
\end{align}
\end{widetext}

\end{appendix}

\bookmarksetup{startatroot}

\bibliographystyle{apsrev}
\bibliography{BibTex_Alexander_Mering}

\end{document}